\documentclass[preprint,pre,longbibliography,eqsecnum]{revtex4-1} 

\usepackage{natbib}
\usepackage{graphicx}
\usepackage{subfigure}
\usepackage{amssymb}
\usepackage{amsfonts}

\usepackage{amsmath}
\usepackage{amsbsy}
\usepackage{mathrsfs} 
\usepackage{color}
\usepackage[colorlinks=true,citecolor=blue]{hyperref}
\usepackage{soul}
\usepackage{lineno}
\usepackage{mymacros}
\usepackage{graphicx}
\begin{document} 

\begin{abstract}
We illuminate effects of surface-charge convection intrinsic to leaky-dielectric electrohydrodynamics by analyzing the symmetric steady state of a circular drop in an external field at arbitrary electric Reynolds number $\mathrm{Re}_E$. 
In formulating the problem, we identify an exact factorization that reduces the number of dimensionless parameters from four --- $\mathrm{Re}_E$ and the conductivity, permittivity and viscosity  ratios --- to two: a modified electric Reynolds number $\widetilde{\mathrm{Re}}$ and a charging parameter $\varpi$. In the case $\varpi<0$, where charge relaxation in the drop phase is slower than in the suspending phase, and, as a consequence, the interface polarizes antiparallel to the external field, we find that above a critical $\widetilde{\mathrm{Re}}$ value the solution exhibits a blowup singularity such that the surface-charge density diverges antisymmetrically with the $-1/3$ power of distance from the equator. We use local analysis to uncover the structure of that blowup singularity, wherein surface charges are convected by a locally induced flow towards the equator where they annihilate. To study the blowup regime, we devise a numerical scheme encoding that local structure where the blowup prefactor is determined by a global charging--annihilation balance. We also employ asymptotic analysis to construct a universal problem governing the blowup solutions in the regime $\widetilde{\mathrm{Re}}\gg1$, far beyond the blowup threshold. In the case $\varpi>0$, where charge relaxation is faster in the drop phase and the interface polarizes parallel to the external field, we numerically observe and asymptotically characterize the formation at large $\widetilde{\mathrm{Re}}$ of stagnant, perfectly conducting surface-charge caps about the drop poles. The cap size grows and the cap voltage decreases  monotonically with increasing conductivity or decreasing permittivity of drop phase relative to suspending phase. The flow in this scenario is nonlinearly driven by electrical shear stresses at the complement of the caps. In both polarization scenarios, the flow at large $\widetilde{\mathrm{Re}}$ scales linearly with the magnitude of the external field, contrasting the familiar quadratic scaling under weak fields. 
\end{abstract}

\title{Equatorial blowup and polar caps in drop electrohydrodynamics}
\author{Gunnar G. Peng$^1$, Rodolfo Brand{\~a}o$^{2}$, Ehud Yariv$^3$, Ory Schnitzer$^1$}
\affiliation{$^1\!$Department of Mathematics, Imperial College London, London SW7 2AZ, UK}
\affiliation{$^2\!$Department of Mechanical and Aerospace Engineering, Princeton University, Princeton, New Jersey 08544, USA}
\affiliation{$^3\!$Department of Mathematics, Technion---Israel Institute of Technology, Haifa 32000, Israel}

\maketitle

\section{Introduction}
\label{sec:introduction}
The problem of a liquid drop suspended in an immiscible liquid and exposed to an external electric field is fundamental to electrohydrodynamics. Taylor's classical theory  \cite{Taylor:66} was founded on his realization that poorly conducting liquids, such as oils, are fundamentally different from perfect dielectrics: however small the conductivity of either fluid, surface charge conducted to the interface by bulk Ohmic currents must keep accumulating there until a steady state is established wherein electric current is continuous across the interface. In turn, the electric field acting on its self-induced surface charge animates a shear-driven flow. Taylor's theory rationalized why in some cases the drop deforms into a flattened oblate-spheroidal shape, rather than an elongated prolate-spheroidal one --- the only possibility predicted by earlier hydrostatic theories. It also formed the basis for the celebrated Taylor--Melcher leaky-dielectric model \cite{Melcher:69},
which is widely employed in electrohydrodynamics modeling   \cite{Saville:97,Fernandez:07,Vlahovska:19,Papageorgiou:19}. 

It is well known that Taylor's theory of drop electrohydrodynamics, wherein the electric and flow fields are, respectively, linear and quadratic in the external field, is restricted to \emph{weak} electric fields. Setting aside the neglect of inertia \cite{Schnitzer:13:LargeRe}, which is appropriate to typical experimental scenarios involving millimetric drops and highly viscous oils, this restriction is associated with two assumptions. The first is that the drop interface is only slightly deformed from its equilibrium spherical shape. Accordingly, the electric and hydrodynamic fields are calculated assuming a spherical interface, with the deformation subsequently derived in a perturbative manner. The extent of deformation is characterized by an electric capillary number representing the ratio of the induced normal stresses to the capillary pressure \cite{Lac:07}. Secondly, convection of surface charge by the induced flow is tacitly neglected, whereby the steady-state charging balance reduces to an interfacial current-continuity condition. The relative strength of surface-charge convection in the original balance is characterized by an electric Reynolds number \cite{Melcher:69}, defined as the ratio of a charge-relaxation time scale, provided by a ratio of permittivity to conductivity, to the convection time scale,  provided by the ratio of the drop size to Taylor's velocity scale. 

Drop electrohydrodynamics is extremely complex beyond the weak-field regime, both in terms of the variety of phenomena and the theoretical challenges posed by deformation and surface-charge convection. To date, the parameter space, spanning five dimensionless numbers --- the electric capillary and Reynolds numbers, alongside the conductivity, permittivity and viscosity ratios --- has only been partially explored. Nonetheless, a multitude of phenomena have been uncovered by experiments and numerical simulations  \cite{Vlahovska:19}. These have been traditionally classified based on whether drop deformation is prolate or oblate under weak fields. Prolate drops become highly elongated under strong fields, in many cases exhibiting breakup above a critical field strength via either formation of lobes followed by pinching or formation of conical ends followed by tip-streaming \cite{Lac:07,Collins:13,Karyappa:14}. 
The strong-field response of oblate drops is even richer, including symmetry breaking to spontaneous ``Quincke rotation'' \cite{Ha:00,Salipante:10,Salipante:13}; the formation of an equatorial belt of vortices  \cite{Dommersnes:13,Ouriemi:14,Ouriemi:15};
 the deformation of the drop into a lens-like shape and its subsequent draining via equatorial streaming \cite{Brosseau:17,Wagoner:20,Wagoner:21}; and breakup via dimpling at the poles \cite{Torza:71,Lac:07,Brosseau:17,Wagoner:20,Wagoner:21}. While the Quincke rotation of sufficiently viscous drops can be understood 
based on the classical theory for rigid particles \cite{Quincke:61,Melcher:69,Jones:84} and  perturbations thereof \cite{He13,Das:17,Das:21}, numerical approaches have hitherto failed \cite{Das:17a,Firouznia:23} --- as further discussed below --- in the case of low-viscosity drops where hysteresis is observed in experiments \cite{Salipante:10}. Perhaps related to this, the equatorial-vortex instability has not yet been found in numerical simulations \cite{Firouznia:23}; while it has been suggested that this experimentally observed instability may merely represent an artefact of the use of insufficiently dilute surface-absorbed particles as tracers \cite{Vlahovska:19}, the phenomenon strikingly resembles surface-electroconvection, a purely electrohydrodynamic instability observed in several setups \cite{Malkus:61,Jolly:70}. 

A seemingly natural approach to theoretically study drop electrohydrodynamics beyond weak fields would involve a linear stability analysis, wherein the ``base'' state is the continuation of Taylor's steady solution to non-zero electric capillary and Reynolds numbers. The fact that this has not yet been attempted may be attributed to the apparent difficulty in calculating the base state in the case of oblate drops --- where instabilities are certainly expected. This difficulty was first discussed by Lanauze \textit{et al.} \cite{Lanauze:15}, who performed initial-value boundary-element simulations assuming axial symmetry and an initially uncharged interface. They found that the steady-state charge-density profile steepens at the drop equator as the field magnitude is increased. Beyond a certain critical magnitude of the applied field (corresponding to a rather moderate electric Reynolds number) a steady state could not be numerically attained; rather, the slope would appear to increase with time until numerical failure. Das and Saintillan \cite{Das:17} also conducted such axisymmetric simulations, encountering similar difficulties: ``the charge discontinuity at the equator is so severe that the boundary element simulations blow up before reaching steady state.'' More recently, Firouznia \textit{et al.} \cite{Firouznia:23} developed a fully three-dimensional spectral method to simulate drop electrohydrodynamics. In the case of low-viscosity drops, they found that ``nonlinear steepening by the flow is sufficiently strong…that increasing the resolution does not avoid ringing [Gibbs phenomenon].'' Furthermore, Firouznia \textit{et al.} \cite{Firouznia:22} observed a similar phenomenon in electrohydrodynamic simulations of a fluid interface subjected to a tangential electric field and a periodic stagnation-like flow. 

The nature of the apparent equatorial singularity in the case of oblate drops has remained unclear, as has the possibility of continuing the steady base state beyond weak fields. Intuitively, the steepening effect is associated with oblate drops polarizing antiparallel to the external field, whereby the field acting on its induced charge drives a flow convecting charges of opposite signs towards the equator \cite{Lanauze:15}. In weak-field theory \cite{Taylor:66}, antiparallel polarization is associated with  slower charge relaxation within the drop phase ---  a necessary  (but not sufficient) condition for oblate deformation. 
Intriguingly, the combination of appreciable surface-charge convection and antiparallel polarization also underlies both the Quincke and surface-electroconvection instabilities. An interplay between singularity formation and those instability mechanisms may accordingly be at the root of the rich electrohydrodynamic phenomena exhibited by oblate-type drops. 

Motivated by this, we shall here analyze the symmetric base state of a drop in an electric field with the purpose of illuminating fundamental effects of surface-charge convection and in particular singularity formation. To that end, we shall neglect shape deformations and address the two-dimensional problem where the drop interface is circular. At zero electric Reynolds number, this problem possesses a unique steady state which constitutes the two-dimensional analog of Taylor's weak-field solution \cite{Feng:02}. Our 
goal will accordingly be to study the continuation of that 
symmetric steady state as the electric Reynolds number is increased to arbitrary values. As such, we will not address here the stability of that steady state nor the nonlinear drop dynamics. In particular, we will not study the transition to the Quincke-rotation states previously calculated in this two-dimensional setup, first numerically  \cite{Feng:02} and later asymptotically for large electric Reynolds number \cite{Yariv:16}. 

Given their fundamental difference, we will separately consider the cases where charge relaxation is slower or faster in the drop phase relative to the suspending phase. This classification can be linked to the polarity of the drop at arbitrary electric Reynolds numbers. In the case where the drop phase relaxes slower we expect to encounter, beyond a critical electric Reynolds number, some \textit{a priori} unknown form of singularity at the drop equator. In contrast to previous studies, we shall seek to understand this singularity formation and even attempt to continue the solution into the singular regime.
 In the case where the drop phase relaxes faster, our focus will be on illuminating the electrohydrodynamics at large electric Reynolds numbers, a regime that has not been explored before. 

This paper is structured as follows. In Sec.~\ref{sec:formulation}, we formulate the exact problem and present several transformations and integral balances that enable significant simplification. Sec.~\ref{sec:preliminary}-\ref{sec:shock_asym} are concerned with the case where charge relaxation is slower in the drop phase. In Sec.~\ref{sec:preliminary}, we present and interpret preliminary numerical results demonstrating the emergence of an equatorial singularity in the steady-state solution. In Sec.~\ref{sec:local}, we carry out a local analysis near the drop equator, uncovering a novel charge-density blowup singularity supported by the leaky-dielectric equations. In Sec.~\ref{sec:global}, we discuss the nature of steady-state solutions exhibiting equatorial blowup and employ  a modified singularity-capturing numerical scheme to study the blowup regime. In Sec.~\ref{sec:shock_asym}, we supplement the latter numerical results by asymptotic analysis in the limit of large electric Reynolds number. The case where charge relaxation is faster in the drop phase is addressed in Sec.~\ref{sec:caps} using a combination of numerical simulations and asymptotic analysis in the large electric-Reynolds-number limit. We discuss our results in Sec.~\ref{sec:discussion} and make concluding remarks in Sec.~\ref{sec:conclusions}, addressing generalizations and further directions.  

\section{Problem formulation}\label{sec:formulation}
\subsection{Dimensionless formulation}
\begin{figure}[t!]
\begin{center}
\includegraphics[scale=0.6]{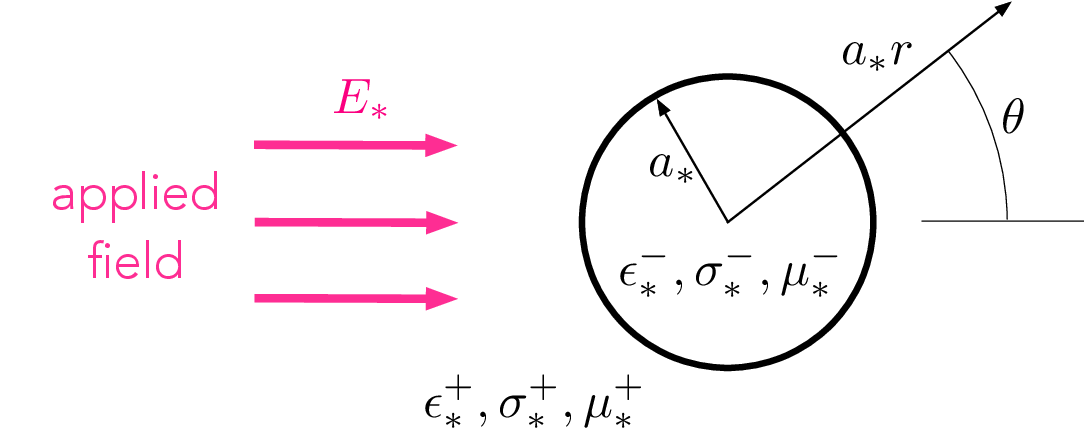}
\caption{Dimensional schematic of a circular leaky-dielectric drop suspended in another leaky-dielectric liquid and subjected to an external electric field.}
\label{fig:sketch}
\end{center}
\end{figure}

We consider a cylindrical leaky-dielectric drop (viscosity ${\mu}_*^-$, electrical conductivity ${\sigma}_*^-$ and permittivity ${\epsilon}_*^-$) that is suspended in a background leaky-dielectric liquid (viscosity $\mu_*^+$, conductivity $\sigma_*^+$ and permittivity $\epsilon_*^+$) of infinite extent. The system is subjected to a uniform electric field of magnitude $E_*$, which is applied perpendicular to the cylinder axis. We only address two-dimensional scenarios, where the resulting electric and flow fields are also perpendicular to the cylinder axis and independent of position along it. Accordingly, we formulate a two-dimensional problem in the cross-sectional plane. We neglect inertia and assume that surface tension is sufficiently strong such that interfacial deformation can be neglected. The cylindrical drop is thus represented by a disk, say of radius $a_*$ --- see Fig.~\ref{fig:sketch}. 

Henceforth, we adopt a dimensionless convention where lengths are normalised by $a_*$, time by $\epsilon_*^+/\sigma_*^+$, potentials by $a_*E_*$, surface-charge density by $\epsilon_*^+ E_*$ and velocities by $v_*=\epsilon_*^+ a_* {E_*}^2/\mu_*^+$. 
The formulation features four dimensionless parameters, namely the three material ratios 
\refstepcounter{equation}
$$
\mathcal{R}=\frac{{\sigma}_*^-}{\sigma_*^+}, \quad \mathcal{S}=\frac{{\epsilon}_*^-}{\epsilon_*^+}, \quad \mathcal{M}=\frac{{\mu}_*^-}{\mu_*^+},
\eqno{(\theequation \mathrm{a},\mathrm{b},\mathrm{c})}
$$
and the electric Reynolds number
\begin{equation}
\mathrm{Re}_E=\frac{{\epsilon_*^+}^2E_*^2}{\mu_*^+\sigma_*^+},
\end{equation}
constituting the ratio of the charge-relaxation time $\epsilon_*^+/\sigma_*^+$ to the convection time $a_*/v_*$. As shown in Fig.~\ref{fig:sketch}, the polar coordinates $(r,\theta)$ are defined with the origin at the drop center and $\theta=0$ in the direction of the applied field. In the suspending fluid, we denote the dimensionless electric potential by $\varphi^+$ and the velocity field by $\bu^+=u^+\be_r+v^+\be_{\theta}$. The corresponding drop-phase fields are denoted by $\varphi^-$ and $\bu^-=u^-\be_r+v^-\be_{\theta}$. We also denote by $\psi^\pm$ the stream functions associated with the velocity fields $\bu^{\pm}$, such that
\refstepcounter{equation}
\label{psi def}
$$
u^{\pm}=\frac{1}{r}\pd{\psi^{\pm}}{\theta}, \quad v^{\pm}=-\pd{\psi^{\pm}}{r}.
\eqno{(\theequation\mathrm{a},\mathrm{b})}
$$
The dimensionless surface-charge density is denoted by $q$. 

The potentials $\varphi^\pm$ are governed by Laplace's equation in their respective domains. The drop potential $\varphi^-$ also satisfies regularity at the origin, while the suspending-fluid potential satisfies the far-field condition
\begin{equation}\label{far}
\varphi^+ = -r\cos\theta +o(1) \quad \text{as} \quad r\to\infty,
\end{equation}
where, without loss of generality, we have fixed the reference level of the potential by eliminating the constant term in its far-field expansion. 
At the interface $r=1$ the potentials satisfy the continuity condition, $\varphi^+={\varphi}^-$, as well as Gauss's law, 
\begin{equation}\label{disp}
q=\mathcal{S}\pd{\varphi^-}{r}-\pd{\varphi^+}{r}. 
\end{equation}
The velocity fields $\bu^{\pm}$ satisfy the continuity and Stokes equations, regularity at the origin and decay at infinity. At $r=1$, they satisfy the impermeability condition, $u^+={u}^-=0$, the continuity condition, $v^+={v}^-$, and the tangential shear-stress balance,
\begin{equation}\label{stress}
\pd{v^+}{r}-v-\mathcal{M}\left(\pd{v^-}{r}-{v}\right)=q\pd{\varphi}{\theta}; 
\end{equation}
hereafter, we omit the $\pm$ superscript when evaluating the fields $\varphi^{\pm}$ and $v^{\pm}$ at $r=1$, where they are continuous.
Given the impermeability condition, the stream functions are uniform at $r=1$; without loss of generality, we set them to zero there. Lastly, the surface-charge density $q$, which couples the electric and flow problems, satisfies the charging condition 
\begin{equation}\label{charging}
\pd{q}{t}+\mathrm{Re}_E\pd{}{\theta}\left(qv\right)=\pd{\varphi^+}{r}-\mathcal{R}\pd{{\varphi}^-}{r},
\end{equation}
wherein the fields are evaluated at $r=1$. We observe that $\mathrm{Re}_E$ provides an indication of the strength of surface-charge convection relative to surface-charge relaxation. 

Following the classical Taylor--Melcher description \citep{Melcher:69}, we do not incorporate surface-charge diffusion into the model. This choice contrasts recent computational studies, where exceedingly weak surface-charge diffusion has been incorporated --- apparently for numerical convenience \cite{Wagoner:20,Wagoner:21}. We shall return to this point in our concluding remarks in Sec.~\ref{sec:conclusions}. 

\subsection{Taylor-symmetric steady state} 
\label{ssec:symmetry}
We shall only consider steady-state solutions. Accordingly, we henceforth omit the time derivative in the charging condition \eqref{charging}. 

It is instructive to briefly consider the case $\mathrm{Re}_E=0$, where the electric problem is decoupled from the flow and an explicit solution is readily  obtained \cite{Feng:02}. One first solves the electric problem governing the potentials $\varphi^{\pm}$, consisting of Laplace's equation in the respective domains, regularity at the origin, the far-field condition \eqref{far}, continuity at the interface and the charging condition \eqref{charging} with the left-hand side set to zero. This gives 
\refstepcounter{equation}
$$
\label{zero Re pot}
 \varphi^-=-\frac{2}{1+\mathcal{R}}r\cos\theta, \quad \varphi^+=-r\cos\theta+\frac{\mathcal{R}-1}{\mathcal{R}+1}\frac{\cos\theta}{r},
\eqno{(\theequation \mathrm{a},\mathrm{b})}
$$
whereby the displacement condition \eqref{disp} provides the surface-charge density as
\begin{equation}\label{zero Re q}
q=\frac{2(\mathcal{R}-\mathcal{S})}{1+\mathcal{R}}\cos\theta.
\end{equation}
The flow problem is then closed and can be solved to give
\begin{equation}\label{zero Re psi}
\psi^{\pm}=\frac{\mathcal{R}-\mathcal{S}}{4(1+\mathcal{M})(1+\mathcal{R})^2}r^{\mp 2}(r^2-1)\sin 2\theta.
\end{equation}
The steady solution given by \eqref{zero Re pot}--\eqref{zero Re psi} constitutes the two-dimensional analog of that found by Taylor for a three-dimensional spherical drop \cite{Taylor:66}. Note that $\varphi^{\pm}$ and $q$ are symmetric about the line connecting the drop poles at $\theta=0,\pi$ and antisymmetric about the ``equatorial'' line $\theta=\pm\pi/2$, while $u^{\pm}$ and $v^{\pm}$ are, respectively, symmetric and antisymmetric about both lines; the stream functions $\psi^{\pm}$ are antisymmetric about both lines.  

For the remainder of this paper, we shall focus on analysing the continuation of the weak-field solution \eqref{zero Re pot}--\eqref{zero Re psi} to $\mathrm{Re}_E>0$. Noting that the aforementioned ``Taylor symmetry'' is compatible with the governing equations at all $\mathrm{Re}_E$, we here seek a  solution satisfying it. Such symmetry implies, at $\theta=0,\pi$, 
\refstepcounter{equation}
$$
\label{sym conditions poles}
\pd{q}{\theta}=0, \quad \pd{\tilde{\varphi}^{\pm}}{\theta}=0,  \quad \pd{\tilde{u}^{\pm}}{\theta}=0,\quad \tilde{v}^{\pm}=0, \eqno{(\theequation \mathrm{a},\mathrm{b},\mathrm{c},\mathrm{d})}
$$
whereas at $\theta=\pm\pi/2$, 
\refstepcounter{equation}
$$
\label{sym conditions equator}
q=0, \quad \tilde{\varphi}^{\pm}=0,  \quad \pd{\tilde{u}^{\pm}}{\theta}=0,\quad \tilde{v}^{\pm}=0.
\eqno{(\theequation \mathrm{a},\mathrm{b},\mathrm{c},\mathrm{d})}
$$
Furthermore, the stream functions satisfy $\psi^{\pm}=\partial^2\psi^{\pm}/\partial\theta^2=0$ at $\theta=0$, $\pm\pi/2$ and $\pi$. 

\subsection{Charging balance}\label{ssec:balance}
Integration of the charging condition \eqref{charging} over the first quadrant $\theta\in(0,\pi/2)$ gives
\begin{equation}\label{gunnar regular before symmetry}
\mathrm{Re}_E\left(\lim_{\theta\nearrow\pi/2}qv-\lim_{\theta\searrow0}qv\right)=\int_{0}^{\pi/2}\left(\pd{\varphi^+}{r}-\mathcal{R}\pd{{\varphi}^-}{r}\right)\mathrm{d}\theta \quad \text{at} \quad r=1,
\end{equation}
where, on physical grounds, we require existence of the integral which represents the net charging of the interface in the first quadrant. For continuous $q$ and $v$, the symmetry conditions (\ref{sym conditions poles}) and (\ref{sym conditions equator}) imply that both limits in \eqref{gunnar regular before symmetry} vanish, whereby 
\begin{equation}\label{gunnar regular}
\int_{0}^{\pi/2}\left(\pd{\varphi^+}{r}-\mathcal{R}\pd{{\varphi}^-}{r}\right)\mathrm{d}\theta=0 \quad \text{at} \quad r=1.
\end{equation}
This integral balance dictates that under Taylor symmetry the net charging of the interface in the first quadrant must vanish at steady state. 

Consider now the possibility that the solution exhibits singular behavior at the equator, specifically such that $q$ diverges as $\theta\to\pm\pi/2$ 
--- a scenario which we shall indeed encounter for $\mathcal{R}<\mathcal{S}$  above a critical value of $\mathrm{Re}_E$.  Then \eqref{gunnar regular} generalizes as
\begin{equation}\label{gunnar singular}
\mathrm{Re}_E\lim_{\theta\to\pi/2}qv=\int_{0}^{\pi/2}\left(\pd{\varphi^+}{r}-\mathcal{R}\pd{{\varphi}^-}{r}\right)\mathrm{d}\theta \quad \text{at} \quad r=1,
\end{equation}
where we have used the symmetry of the surface current $qv$ about the equatorial line to replace the one-sided limit $\theta\nearrow\pi/2$ appearing in \eqref{gunnar regular before symmetry} with the two-sided limit $\theta\to\pi/2$. 
Given that $q$ and $v$ are antisymmetric about the equatorial line, the double-sided limit of the surface current implies ``annihilation'' or ``creation'' of positive and negative charges there --- respectively, depending on whether the surface flow is locally towards or away from the equator. Thus, we interpret condition \eqref{gunnar singular} as a generalized Ohmic-charging balance which accounts for such possibilities. (We shall later see that the form of singularity that the solution can develop is associated with annihilation.)

\subsection{``Factorized'' formulation based on reflection}
\label{ssec:reduced}
Prior to any (numerical or asymptotic) analysis, the problem governing the Taylor-symmetric steady state can be significantly simplified. We first define the ``external'' harmonic potential 
\begin{equation}
\Phi^+(r,\theta)= \left(\frac{1}{r}-r\right)\cos\theta,
\end{equation}
which satisfies the far-field approach to the external field as in  \eqref{far} and vanishes at $r=1$. We can then derive the reflection relations
\refstepcounter{equation}
$$
\label{reflections}
\pd{}{r}\left({\varphi}^+-\Phi^+\right)=-\pd{{\varphi}^-}{r}, \quad \pd{v^+}{r}-v=-\left(\pd{{v}^-}{r}-v\right) \quad \text{at} \quad r=1.
\eqno{(\theequation\mathrm{a},\mathrm{b})}
$$
The first reflection relation,  (\ref{reflections}a), follows from the observation that 
\begin{equation}\label{pot reflection}
\varphi^+(r,\theta)-\Phi^+(r,\theta)={\varphi}^-(1/r,\theta) \quad \text{for} \quad r\geq 1.
\end{equation} 
Indeed, the two sides of this relation satisfy Laplace's equation for $r>1$, are equal at $r=1$ and decay as $r\to\infty$. [Note that ${\varphi}^-(1/r,\theta)$ vanishes as $r\to\infty$ since the regularity of $\varphi^-(r,\theta)$ at the origin in conjunction with the Taylor symmetry implies that $\lim_{r\searrow0}\varphi^-(r,\theta)=0$.] Relation \eqref{pot reflection} thus follows from uniqueness of solutions to Laplace's equation. The second reflection relation, (\ref{reflections}b), similarly follows from the observation that 
\begin{equation}\label{psi reflection}
\psi^+(r,\theta)=-r^2\psi^-(1/r,\theta)  \quad \text{for} \quad r\geq 1.
\end{equation}
In this case, the two sides of the relation are biharmonic for $r>1$, have zero tangential derivative and equal radial derivative at $r=1$ and have decaying gradients as $r\to\infty$. [Note that the regularity of $\psi^-(r,\theta)$ at the origin in conjunction with the Taylor symmetry implies that $\psi^-(r,\theta)\sim \text{const.} \times r^2\sin 2\theta$ as $r\to0$, so that $r^2\psi^-(1/r,\theta)\sim \text{const.} \times \sin 2\theta$ as $r\to\infty$.] Relation \eqref{psi reflection} thus follows from uniqueness of solutions to the biharmonic equation. 

It is clear from \eqref{disp}--\eqref{charging} that the reflection relations \eqref{reflections} allow us to reduce the problem to either the interior or exterior domains. Furthermore, they reveal a non-trivial factorization of the fields that effectively reduces the number of parameters in the problem \emph{from four to two}. The two independent parameters can be chosen as the ``modified electric Reynolds number''
\begin{equation}\label{Re tilde}
\widetilde{\mathrm{Re}}=\frac{\mathrm{Re}_E}{(1+\mathcal{M})(1+\mathcal{S})}
\end{equation}
and the ``charging parameter'' 
\begin{equation}
\label{varpi}
\varpi=\frac{1+\mathcal{R}}{1+\mathcal{S}}-1,
\end{equation}
whose range is evidently $\varpi>-1$. Indeed, upon defining the rescaled fields $\tilde{\varphi}^{\pm}$ and $\tilde{\bu}^{\pm}$ via
\refstepcounter{equation}
$$
\label{tilde def}
\varphi^+=\Phi^++\frac{1}{1+\mathcal{S}}\tilde{\varphi}^+, \quad \varphi^-=\frac{1}{1+\mathcal{S}}\tilde{\varphi}^-, \quad \bu^{\pm}=\frac{1}{(1+\mathcal{M})(1+\mathcal{S})}\tilde{\bu}^{\pm},
\eqno{(\theequation\mathrm{a},\mathrm{b},\mathrm{c})}
$$
conditions \eqref{disp}--\eqref{charging} can be written, say in terms of the interior rescaled fields, 
\refstepcounter{equation}
$$
\label{factored equations}
q=\pd{\tilde{\varphi}^-}{r}+2\cos\theta, \quad \pd{\tilde{v}^-}{r}-\tilde{v}=-q\pd{\tilde{\varphi}}{\theta}, \quad \widetilde{\mathrm{Re}}\pd{}{\theta}\left(q\tilde{v}\right)=-(1+\varpi)\pd{\tilde{\varphi}^-}{r}-2\cos\theta,
\eqno{(\theequation\mathrm{a},\mathrm{b},\mathrm{c})}
$$
respectively. We also define the stream functions $\tilde{\psi}^{\pm}$, corresponding to the rescaled velocity fields $\tilde{\bu}^{\pm}$ as in \eqref{psi def}. The problem therefore reduces to obtaining a harmonic potential $\tilde{\varphi}^-$ and a biharmonic stream function $\tilde{\psi}^-$ that possess Taylor symmetry (see Sec.~\ref{ssec:symmetry}) and satisfy conditions \eqref{factored equations} together with regularity conditions at $r=0$. Once this problem has been solved, the suspending-fluid potential $\tilde{\varphi}^+$ and stream function $\tilde{\psi}^+$ are readily obtained from the reflection relations \eqref{pot reflection} and \eqref{psi reflection}, respectively. (Of course, the problem could alternatively be formulated in terms of the exterior fields.) 

Using the displacement condition (\ref{factored equations}a), we  eliminate the rescaled potential from the charging condition (\ref{factored equations}c). This gives the modified charging condition
\begin{equation}
\widetilde{\mathrm{Re}}\pd{}{\theta}\left(q\tilde{v}\right)+(1+\varpi)q=2\varpi\cos\theta,
\label{charging in q}
\end{equation}
which we shall employ instead of (\ref{factored equations}c). Motivated by the simplified form of the modified charging condition \eqref{charging in q}, we repeat the derivation of the generalized charging balance \eqref{gunnar singular}, which accounts for the possibility of singular behavior at the equator. Thus, we integrate  \eqref{charging in q} over the first quadrant and use the symmetry condition (\ref{sym conditions poles}d) at $\theta=0$, as well as the symmetry of $q\tilde{v}$ about the equatorial line, to find
\begin{equation}\label{q quad singular}
\widetilde{\mathrm{Re}}\lim_{\theta\to\pi/2} q\tilde{v}+(1+\varpi)\mathcal{Q}=2\varpi,
\end{equation} 
where we define the quadrant charge
\begin{equation}\label{Q def}
\mathcal{Q}=\int_0^{\pi/2}q\,\mathrm{d}\theta.
\end{equation}
As in \eqref{gunnar singular}, the limit term in \eqref{q quad singular} represents charge annihilation or creation at the equator depending on whether the surface flow is locally towards or away from the equator. Similarly to the Ohmic-charging integral in \eqref{gunnar singular}, we require on physical grounds that the surface-charge integral in \eqref{Q def} exists. 

If $q$ and $\tilde{v}$ are continuous, the symmetry conditions \eqref{sym conditions equator} imply that $\lim_{\theta\to\pi/2} q\tilde{v}=0$. Remarkably, \eqref{q quad singular} then reduces to an explicit formula for the quadrant charge, 
\begin{equation}\label{q quad regular}
\mathcal{Q}=\frac{2\varpi}{1+\varpi},
\end{equation} 
which is independent of $\widetilde{\mathrm{Re}}$. This result constitutes the factorized counterpart of the charging balance \eqref{gunnar regular}. 
\begin{figure}[b!]
\begin{center}
\includegraphics[scale=0.6]{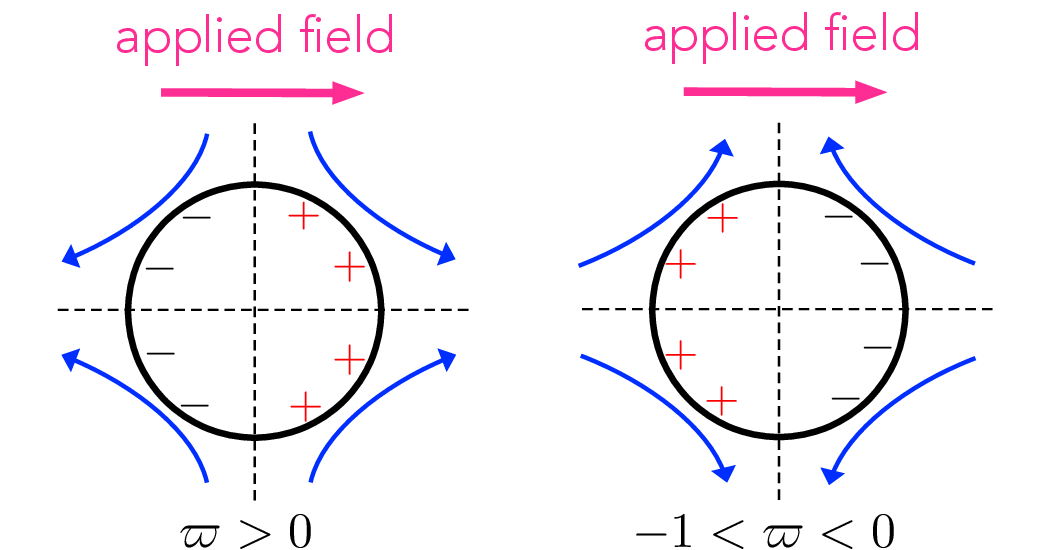}
\caption{Schematic of the parallel- (left) and antiparallel- (right) polarization scenarios. The parameter $\varpi$ is defined in \eqref{varpi} (or \eqref{varpi difference}).}
\label{fig:scenarios}
\end{center}
\end{figure}

\subsection{Polarity}
\label{ssec:polarity}
In calculating the Taylor-symmetric steady state, it is useful to distinguish between the cases of ``parallel polarization,'' $\varpi>0$, and ``antiparallel polarization,'' $-1<\varpi<0$. This terminology reflects the fact that for $\varpi>0$, 
\refstepcounter{equation}
\label{polarity result}
$$
\label{q sign positive}
q>0 \quad \text{for} \quad 0<\theta<\pi/2
\eqno{(\theequation\mathrm{a})}
$$
while, for $\varpi<0$, 
$$
\label{q sign negative}
q<0 \quad \text{for} \quad 0<\theta<\pi/2.
\eqno{(\theequation\mathrm{b})}
$$
In both cases, the corresponding signs of $q$ in the other quadrants follow from that symmetry, see Fig.~\ref{fig:scenarios}. Note that, in particular, the quadrant charge $\mathcal{Q}\gtrless0$ for $\varpi\gtrless0$. The polarity result \eqref{polarity result} is shown in Appendix \ref{app:polarity} to follow from the problem formulation and Taylor symmetry for all $\widetilde{\mathrm{Re}}$ --- even when accounting for the possibility of charge annihilation at the equator. 

By rewriting \eqref{varpi} in the alternative form
\begin{equation}\label{varpi difference}
\varpi=\frac{\mathcal{R}-\mathcal{S}}{1+\mathcal{S}},
\end{equation} 
we see that parallel polarization corresponds to the case $\mathcal{R}>\mathcal{S}$, where the drop phase relaxes faster than the suspending phase, while antiparallel polarization corresponds to the case $\mathcal{R}<\mathcal{S}$, where the suspending phase relaxes faster. In particular, the extreme limits $\varpi\to\infty$ and $\varpi\searrow-1$ correspond to a drop of infinite conductivity and infinite permittivity, respectively. 

For $\varpi=0$, the drop and suspending phases relax equally fast. In that borderline case, the problem for all $\widetilde{\mathrm{Re}}$ possesses a simple solution where the charge density $q$ and the velocity fields $\tilde{\mathbf{u}}^{\pm}$ vanish identically, with the interior potential accordingly given by $\tilde{\varphi}^-=-2r\cos\theta$, representing a uniform field. 

In particular, the above results on polarity are satisfied for the weak-field solution \eqref{zero Re pot}--\eqref{zero Re psi}, which in factorized form reduces to
\refstepcounter{equation}
$$
\label{zero Re solution tilde}
\tilde{\psi}^-=\frac{\varpi }{4(1+\varpi)^2}r^2(r^2-1)\sin 2\theta, \quad q=\frac{2\varpi}{1+\varpi}\cos\theta, \quad \tilde{\varphi}^-=-\frac{2}{1+\varpi}r\cos\theta.
\eqno{(\theequation \mathrm{a},\mathrm{b},\mathrm{c})}
$$
We note that the flow corresponding to (\ref{zero Re solution tilde}a) is directed from the equator to the poles for parallel polarization, and from the poles towards the equator for antiparallel polarization. Our numerical and asymptotic solutions in the following sections suggest that these flow directions persist for arbitrary $\widetilde{\mathrm{Re}}$, though we shall neither prove nor assume this. For convenience, we depict these flow directions in Fig.~\ref{fig:scenarios}. 

\section{From equatorial steepening to blowup}
\label{sec:preliminary}
From this section until Sec.~\ref{sec:shock_asym}, we shall be focusing upon the case $-1<\varpi<0$, where the charge density is polarized antiparallel to the external field [cf.~\eqref{q sign negative}]. 
\begin{figure}[t!]
\begin{center}
\includegraphics[scale=0.6,trim={2.5cm 1cm 4.5cm 0}]{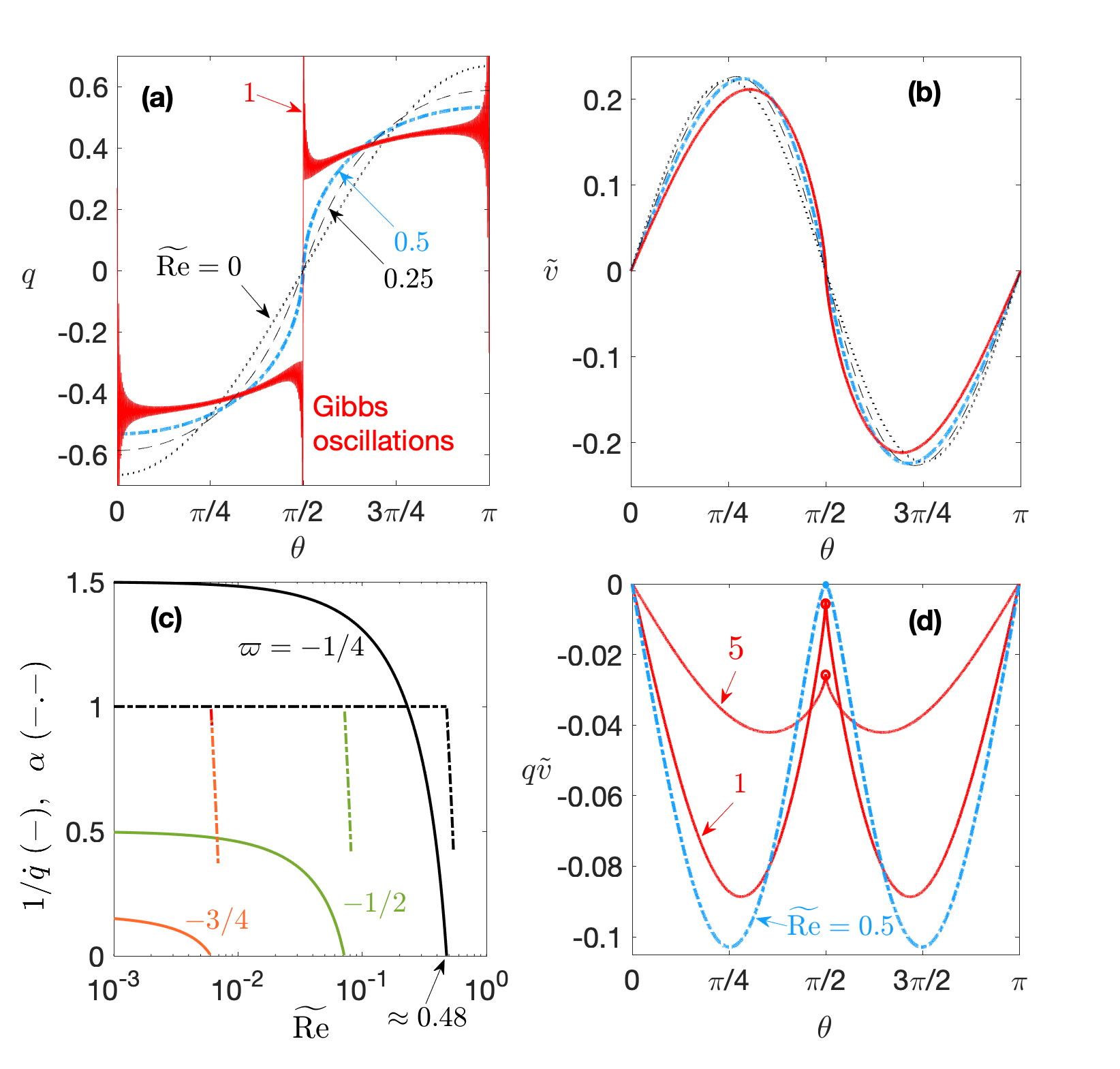}
\caption{Preliminary numerical solutions obtained using the straightforward Fourier-series scheme (see Sec.~\ref{sec:preliminary}). (a) Surface-charge density $q$ and (b) azimuthal surface velocity $\tilde{v}$ for $\varpi=-1/4$ and the indicated values of $\widetilde{\mathrm{Re}}$. (c) Reciprocal of the slope $\dot{q}=(\mathrm{d}q/\mathrm{d}\theta)_{\theta=\pi/2}$ and the exponent $\alpha$ [see \eqref{sublinear}]  as a function of $\widetilde{\mathrm{Re}}$ and for the indicated values of $\varpi$. (d) Surface current $q\tilde{v}$ for $\varpi=-1/4$ and the indicated values of $\widetilde{\mathrm{Re}}$.}
\label{fig:omnegative_prelim}
\end{center}
\end{figure}
We begin in this section by discussing preliminary numerical solutions obtained using a straightforward Fourier-series scheme (described in Appendix \ref{app:numerics_fourier}). We also interpret these solutions with the help of exact and scaling relations.  

Fig.~\ref{fig:omnegative_prelim}a,b show profiles of the charge density $q$ and surface velocity $\tilde{v}$ in the case $\varpi=-1/4$, for $\widetilde{\mathrm{Re}}=0.25$, $0.5$ and $1$; also depicted is the weak-field solution \eqref{zero Re solution tilde} for $\widetilde{\mathrm{Re}}=0$. In these plots, the azimuthal range is confined to $0<\theta<\pi$, corresponding to the drop ``upper half''; alluding to the Taylor symmetry stipulated in subsection \ref{ssec:symmetry}, the extension to the ``lower half'' follows from the symmetries about the pole line whereas the symmetries about the equatorial line are evident in the plots. We note that the sign of $q$ over the front and back halves of the drop are as predicted by the polarity result \eqref{q sign negative}, and that the surface velocity is directed from the poles to equator in all of the cases shown. 

\subsection{Equatorial steepening}
Going from $\widetilde{\mathrm{Re}}=0$ to $0.25$, surface-charge convection is seen to steepen the gradients of $q$ and $\tilde{v}$ at the equator --- henceforth denoted $\dot{q}=(\mathrm{d}q/\mathrm{d}\theta)_{\theta=\pi/2}$ and $\dot{\tilde{v}}=(\mathrm{d}\tilde{v}/\mathrm{d}\theta)_{\theta=\pi/2}$. These two slopes can be explicitly related. Indeed, supposing smooth solutions possessing Taylor symmetry, a local balance between all three terms in the charging equation \eqref{charging in q} readily gives 
\begin{equation}
\dot{q}=-\frac{2\varpi}{1+\varpi+2\dot{\tilde{v}}\widetilde{\mathrm{Re}}},
\label{q slope}
\end{equation}
showing that $\dot{q}$ diverges as $-\dot{\tilde{v}}\widetilde{\mathrm{Re}}$ increases towards the positive value $(1+\varpi)/2$. The divergence of $\dot{q}$ is numerically demonstrated in Fig.~\ref{fig:omnegative_prelim}c for several negative values of $\varpi$. [Instead of directly computing the diverging quantity $\dot{q}$, we compute $\dot{\tilde{v}}$ and then extract $\dot{q}$ from \eqref{q slope}.] We see in that figure that the critical value of $\widetilde{\mathrm{Re}}$ at which $\dot{q}$ diverges increases with $\varpi$. In particular, for $\varpi=-1/4$ the slope diverges at $\widetilde{\mathrm{Re}}\approx 0.48$. Clearly, to continue the steady solution beyond this divergence we must accept the possibility of non-smoothness at the equator. 

\subsection{Transition}
\label{ssec:transition}
Consider the solution depicted in Fig.~\ref{fig:omnegative_prelim}a,b for $\varpi=-1/4$ and $\widetilde{\mathrm{Re}}=0.5$, just above the critical value at which $\dot{q}$ diverges. The $q$ profile is continuous but has infinite slope at the equator, while $\tilde{v}$ appears to remain linear there. 

To investigate this behavior, we postulate the sublinear scaling
\begin{equation}
q\sim - \text{const.} \times |\pi/2-\theta|^{\alpha}\mathrm{sgn}(\pi/2-\theta) \quad \text{as} \quad  \pi/2\to\theta,
\label{sublinear}
\end{equation} 
wherein $0<\alpha<1$. [The scaling \eqref{sublinear} also holds, with $\alpha=1$, for $\widetilde{\mathrm{Re}}$ below the critical value at which $\dot{q}$ diverges.] Given Gauss's law (\ref{factored equations}a) and the isotropy of Laplace's equation, \eqref{sublinear} suggests locally induced potential variations of order $\rho^{1+\alpha}$ as the distance $\rho$ from the equator vanishes (i.e., electric field of order $\rho^{\alpha}$), negligible relative to the $\mathcal{O}(\rho)$ potential variations expected from global considerations (i.e., by locally extrapolating a smooth Taylor-symmetric potential). In turn, the electric stress on the right-hand side of the tangential-stress balance (\ref{factored equations}b) scales as $q$, i.e., as $\rho^{\alpha}$; given the isotropy of the Stokes equations, that balance implies a locally induced velocity field of order $\rho^{1+\alpha}$, negligible relative to the $\mathcal{O}(\rho)$ local extrapolation of the global Taylor-symmetric velocity field. In particular, the surface velocity behaves linearly,  
$\tilde{v}\sim -(\pi/2-\theta)\dot{\tilde{v}}$, in accordance with the numerical behavior noted above. 

Upon substituting the local scalings of $q$ and $\tilde{v}$ into the charging equation \eqref{charging in q}, we find that the leading-order balance between the first two terms gives the exponent $\alpha$ as 
\begin{equation}\label{exponent}
\alpha=-\frac{1+\varpi+\dot{\tilde{v}}\widetilde{\mathrm{Re}}}{\dot{\tilde{v}}\widetilde{\mathrm{Re}}}. 
\end{equation}
According to this result, $\alpha$ decreases from unity to zero as $-\dot{\tilde{v}}\widetilde{\mathrm{Re}}$ doubles from the aforementioned critical value $(1+\varpi)/2$ to $1+\varpi$. Fig.~\ref{fig:omnegative_prelim}c depicts computed values of $\alpha$ as a function of $\widetilde{\mathrm{Re}}$, for several negative values of $\varpi$; $\alpha$ is seen to rapidly decrease over a narrow $\widetilde{\mathrm{Re}}$ interval. (As  discussed in Appendix \ref{app:numerics_singular}, precise computation of $\alpha$ using the straightforward Fourier scheme becomes impractical for sufficiently small $\alpha$, typically below $0.4$.)  

\subsection{Blowup}
\label{ssec:blowup}
With the drop polarized antiparallel to the external field, the field acting on its own surface charge tends to drive an electrohydrodyamic flow that convects opposite charges towards the equator. As we have seen, both the steepening of $q$ in the initial ``smooth'' regime and the subsequent decrease in the exponent $\alpha$ in the following ``transition'' regime are determined by that drop-scale flow, specifically through the scaled equatorial slope $\dot{\tilde{v}}\widetilde{\mathrm{Re}}$. In other words, these phenomena are globally driven; conversely, the local behavior of $q$ has no global ramifications in these regimes. 
But the transition regime terminates as $-\dot{\tilde{v}}\widetilde{\mathrm{Re}}$ increases to $1+\varpi$; in that limit, $\alpha\searrow0$ and accordingly the locally induced potential and velocity fields become as large as anticipated based on global considerations. The  local asymptotic structure must accordingly adapt. 

Numerically, the straightforward Fourier-series scheme of Appendix \ref{app:numerics_fourier} fails beyond the transition regime, in the sense that the computed $q$ profiles exhibit Gibbs oscillations --- see Fig.~\ref{fig:omnegative_prelim}a for $\widetilde{\mathrm{Re}}=1$. Nonetheless, the computed $\tilde{v}$ profiles do not exhibit similar oscillations for the same parameters; see Fig.~\ref{fig:omnegative_prelim}b for $\widetilde{\mathrm{Re}}=1$, and note that $\tilde{v}$ appears to continuously vanish as $\theta\to\pi/2$, though not linearly. The surface current $q\tilde{v}$ is similarly computable, with example profiles shown in  Fig.~\ref{fig:omnegative_prelim}d in the case $\varpi=-1/4$, for $\widetilde{\mathrm{Re}}=0.5$, $1$ and $5$. For $\widetilde{\mathrm{Re}}=0.5$, $q\tilde{v}$ continuously vanishes as $\theta\to\pi/2$ as expected in the transition regime. In contrast, for $\widetilde{\mathrm{Re}}=1$ and $5$, $q\tilde{v}$ appears to approach a finite non-zero value in that limit  ---  indicating charge annihilation at the equator [see discussion following \eqref{gunnar singular}, noting that the surface flow here is towards the equator]. Clearly, the apparent local behaviors of $\tilde{v}$ and $q\tilde{v}$ together suggest that $q$ diverges towards the equator! 

We are thus led to consider the possibility where $q$ behaves according to \eqref{sublinear}, but now with $\alpha<0$, corresponding to the surface-charge density diverging antisymmetrically about the equator. In that case, the locally induced fields become dominant near the equator. Indeed, the locally induced potential, of  order $\rho^{1+\alpha}$, is now asymptotically larger than the $\mathcal{O}(\rho)$ potential variations anticipated based on global considerations. As a consequence, the electric stress on the right-hand side of (\ref{factored equations}b) now scales as the product of $q$ and the locally induced potential gradient, i.e., like $\rho^{2\alpha}$. In turn, the locally induced velocity field scales like $\rho^{1+2\alpha}$, which is asymptotically larger than the $\mathcal{O}(\rho)$ velocity anticipated based on global considerations. Given these scalings, surface-charge convection locally dominates charge relaxation in \eqref{charging in q}; as a consequence, the surface current $q\tilde{v}$ approaches a finite non-zero value as $\theta\to\pi/2$ and is accordingly of order unity in that limit. It follows that $\alpha=-1/3$. 

The above scaling analysis, together with the preliminary numerical results, suggests that the transition regime is followed by a blowup regime wherein the charge density is of order $\rho^{-1/3}$ and the velocity field is of order $\rho^{1/3}$ as the distance from the equator $\rho$ vanishes. The following three sections focus on studying this blowup regime. In Sec.~\ref{sec:local}, we carry out a local analysis of the blowup singularity. In Sec.~\ref{sec:global}, we show how the blowup strength is globally determined, and explore the onset and evolution of the blowup singularity in parameter space using a modified singularity-capturing numerical scheme. In Sec.~\ref{sec:shock_asym}, we asymptotically analyze the blowup regime in the limit $\widetilde{\mathrm{Re}}\to\infty$.  

\section{Local analysis of blowup singularity} 
\label{sec:local}
\begin{figure}[t!]
\begin{center}
\includegraphics[scale=0.4,trim={1cm 1cm 0cm 0cm}]{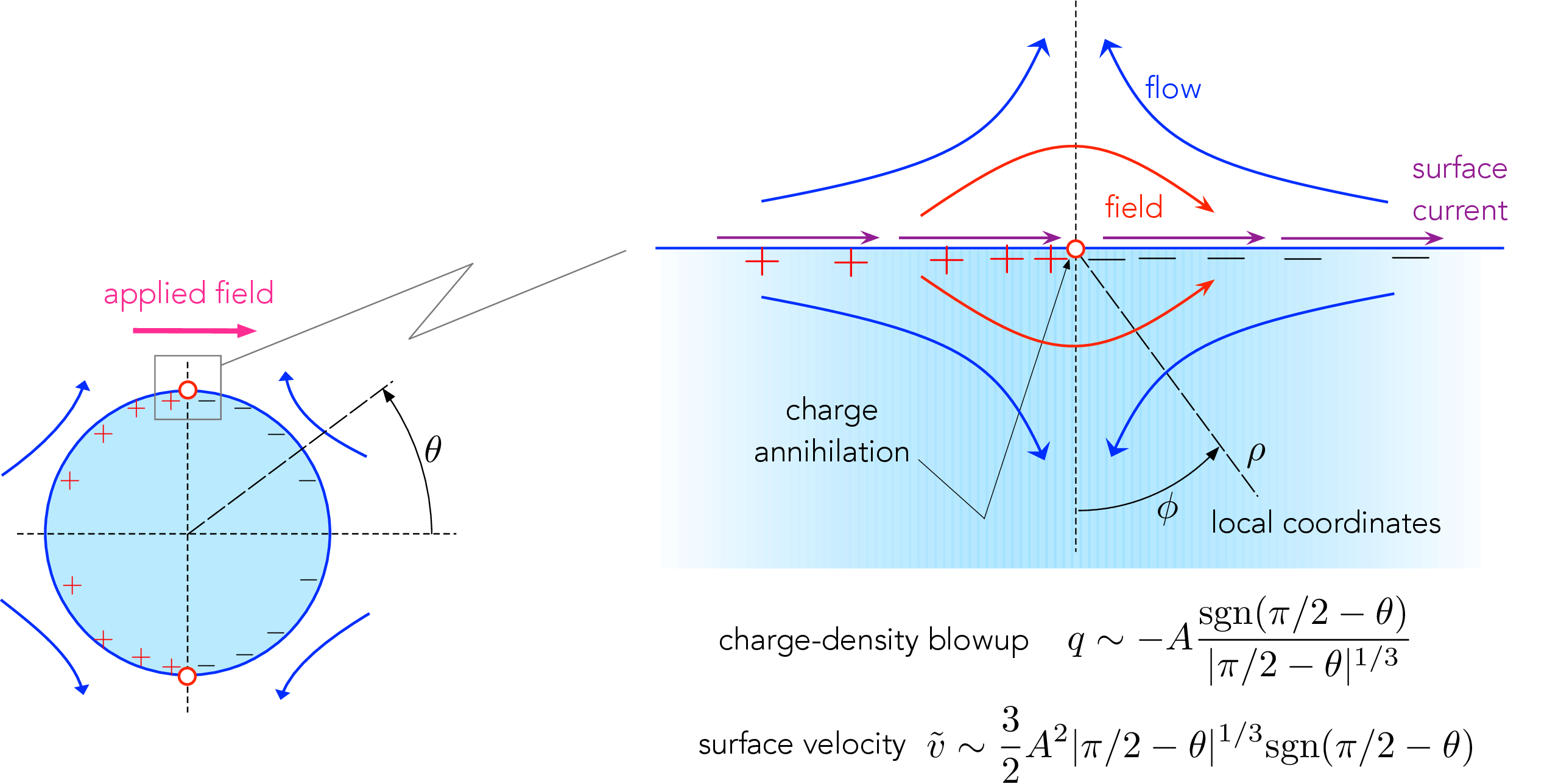}
\caption{Schematic of the local blowup singularity structure identified in Sec.~\ref{sec:local}.}
\label{fig:local_sketch}
\end{center}
\end{figure}

We perform a local analysis approaching the equatorial point $(r,\theta)=(1,\pi/2)$. [The local behavior approaching the other equatorial point $(r,\theta)=(1,-\pi/2)$ follows from Taylor symmetry.] 
Consistently with the above scaling analysis, let ($\rho,\phi)$ be polar coordinates such that $\rho$ is the distance from that point and $\phi$ the angle measured counter-clockwise from $-\be_{r}$ at that point (see Fig.~\ref{fig:local_sketch}).  To leading order as $\rho\searrow0$, 
\begin{equation}
1-r= \rho\cos\phi, \quad
\pi/2-\theta = \rho\sin\phi,\label{def rho phi}
\end{equation}
with the locally flat interface at $\phi=\pm\pi/2$. In what follows, we seek leading-order approximations as $\rho\searrow0$ for the surface-charge density $q$, drop-phase potential $\tilde{\varphi}^-$ and drop-phase stream function $\tilde{\psi}^-$. 

Motivated by the scaling analysis in Sec.~\ref{ssec:blowup}, we derive a local blowup structure where $q$, $\tilde{\varphi}^-$ and $\tilde{\bu}^{-}$ are of order $\rho^{-1/3}$, $\rho^{2/3}$ and $\rho^{1/3}$, respectively. Since 
 $\tilde{\psi}^-$ vanishes at the interface, it is of order $\rho^{4/3}$. Owing to the Taylor symmetry, the fields $q$, $\tilde{\varphi}^-$ and $\tilde{\psi}^-$ are odd in $\phi$. It is accordingly convenient to temporarily restrict our attention to $0<\phi<\pi/2$ (within the first quadrant of the disk). In particular, let $q= - A\rho^{-1/3}$ at $\phi=\pi/2$, where $A$ is a constant. The associated potential $\tilde\varphi^-$, which satisfies Laplace's equation and Gauss's law [cf~(\ref{factored equations}{a})]
\begin{equation}\label{local potential}
\frac{1}{\rho}\pd{\tilde{\varphi}^-}{\phi}= -A\rho^{-1/3} \quad \text{at} \quad  \phi=\pi/2,
\end{equation}
is given by 
\begin{equation}
\tilde\varphi^-=-3A\rho^{2/3}\sin\frac{2\phi}{3}.
\end{equation}
The solution of the biharmonic equation that is proportional to $\rho^{4/3}$, satisfies the impermeability condition, $\tilde\psi^-=0$ at $\phi=\pi/2$, and is odd in $\phi$ is (see \citep{Leal:book})
\begin{equation}
\tilde\psi^- = B\rho^{4/3} \left(\sin\frac{4\phi}{3}-\sin\frac{2\phi}{3}\right),
\label{local psi}
\end{equation} 
where $B$ is a constant. By substituting \eqref{local psi} into the shear condition [cf~(\ref{factored equations}{b})]
\begin{equation}\label{local shear}
\frac{1}{\rho^2}\pd{^2\tilde\psi^-}{\phi^2} = \frac{1}{\rho}\pd{\tilde\varphi^-}{\rho}\pd{\tilde\varphi^-}{\phi}
\quad \text{at} \quad \phi= \pi/2,
\end{equation}
we find $B=3A^2/2$. Since the product $q\tilde{v}$ is independent of $\rho$, the charging balance \eqref{charging in q} is trivially satisfied at order $1/\rho$. (The second and third terms in that balance, which correspond to Ohmic charging from the bulk, are negligible at that  order.) 

We conclude that the problem formulation is compatible with the local blowup behavior
\refstepcounter{equation}
$$
q\sim -A(\pi/2-\theta)^{-1/3}, \quad \tilde{v}\sim \frac{3}{2}A^2(\pi/2-\theta)^{1/3} \quad \text{as} \quad \theta\nearrow\pi/2,
\eqno{(\theequation \mathrm{a},\mathrm{b})}
\label{blow up}
$$
with the limits formed at the other side of the equator, as $\theta\searrow\pi/2$, being opposite in sign. Combining these one-sided behaviors, the two-sided local limit of the surface current is found as
\begin{equation}\label{qv lim}
\lim_{\theta\to\pi/2} q\tilde{v} = -\frac{3}{2}A^3.
\end{equation}
Incidentally, we observe that the corresponding local asymptotic behavior of the grouping $q^2/\tilde{v}$ is independent of $A$,
and accordingly of both $\widetilde{\mathrm{Re}}$ and $\varpi$:
\begin{equation}\label{local collapse}
{q^2}/{\tilde v} \sim \frac{2}{3}(\pi/2-\theta)^{-1}\quad \text{as} \quad \theta\nearrow\pi/2.
\end{equation}

The blowup prefactor $A$ cannot be determined from local considerations alone. We note, however, that while the sign of $A$ is arbitrary in the local analysis, the (global) polarity result \eqref{q sign negative} implies (in the present antiparallel-polarization scenario $-1<\varpi<0$) that $A\ge0$. [We shall confirm in Sec.~\ref{sec:caps} that an equatorial singularity does not form in the parallel-polarization scenario $\varpi>0$, essentially because the surface velocity is then directed away from the equator in contradiction to (\ref{blow up}b).] 

The local singularity structure that we have uncovered above is schematically depicted in Fig.~\ref{fig:local_sketch}. Physically, surface charges of opposite sign are convected towards the equator, where they annihilate. Ohmic charging from the bulk is locally negligible. 

\section{Blowup regime}\label{sec:global}
\subsection{Charging--annihilation balance}
Substituting the surface-current limit \eqref{qv lim} into the integral charging balance \eqref{q quad singular}, we find an exact relation between the singularity prefactor $A$ (a local property) and the quadrant charge $\mathcal{Q}$ (a global quantity),
\begin{equation}
\label{charging annal}
-\frac{3}{2}\widetilde{\mathrm{Re}}\,A^3+(1+\varpi)\mathcal{Q}-2\varpi=0.
\end{equation}
This relation represents a balance between annihilation of charge at the equator (first term) and  charging of the interface via bulk Ohmic currents (last two terms). 

For $A=0$, the Ohmic charging terms in \eqref{charging annal} balance and $\mathcal{Q}$ is given by its pre-blowup value \eqref{q quad regular}. For $A\ne0$,  
$\mathcal{Q}$ is smaller in absolute magnitude. With $A\ge0$ in the present antiparallel-polarization scenario, it is convenient to rewrite \eqref{charging annal} in the form
\begin{equation}
A=\frac{2^{1/3}}{3^{1/3}\widetilde{\mathrm{Re}}^{1/3}}\left[-2\varpi+(1+\varpi)\mathcal Q\right]^{1/3}. \label{A in Q}
\end{equation}

\begin{figure}[t!]
\begin{center}
\includegraphics[scale=0.45,trim={4cm 1cm 4cm 0}]{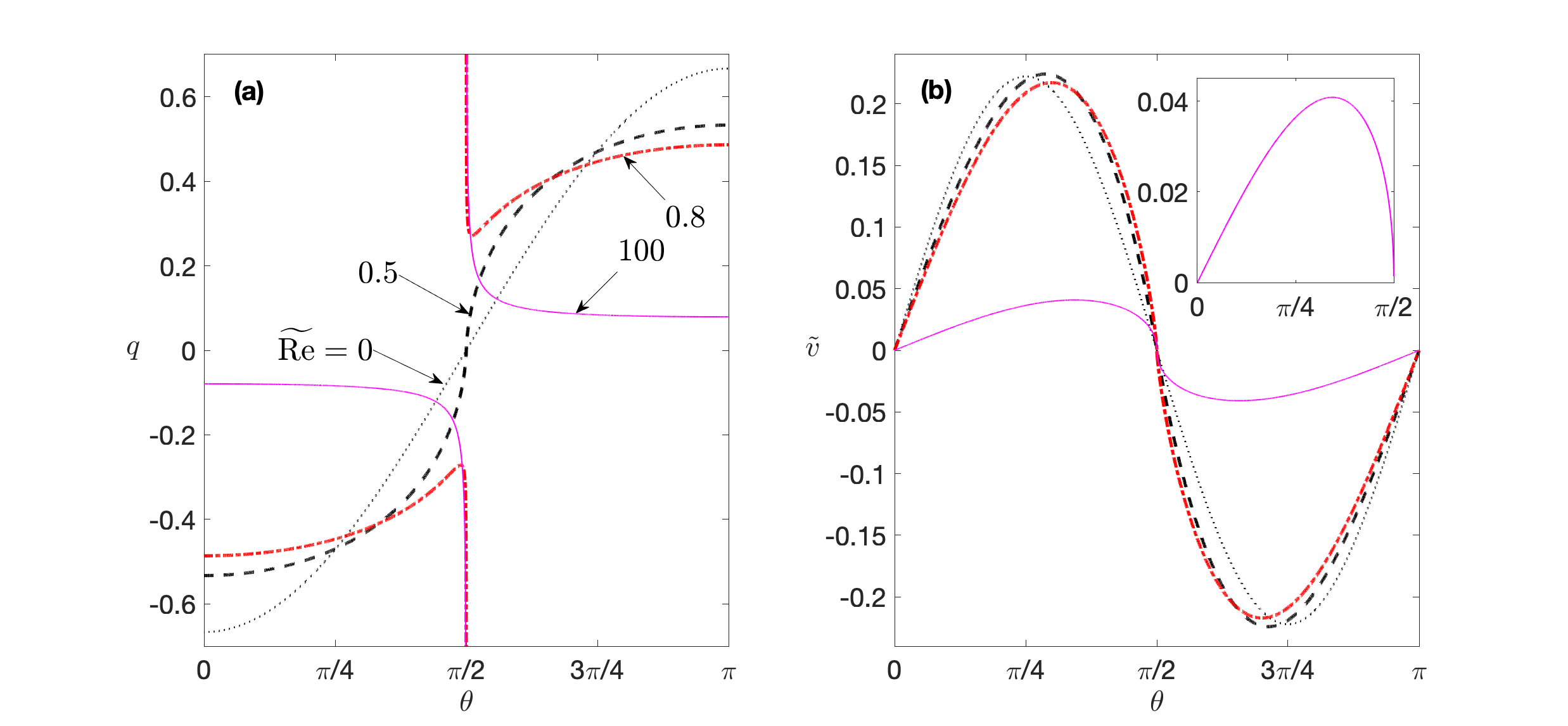}
\caption{Steady state solutions exhibiting equatorial blowup for sufficiently large $\widetilde{\mathrm{Re}}$, numerically calculated using the modified singularity-capturing Fourier-series scheme (see Sec.~\ref{ssec:singularsolutions}). (a) Surface-charge density $q$ and (b) surface velocity $\tilde{v}$ for $\varpi=-1/4$ and several values of $\widetilde{\mathrm{Re}}$: 0 (dotted), 0.5 (dashed), 0.8 (dash-dotted) and 100 (solid); for the latter value, the inset of (b) depicts the surface velocity more clearly.}
\label{fig:omnegative_singular}
\end{center}
\end{figure}
\begin{figure}[t!]
\begin{center}
\includegraphics[scale=0.42,trim={5.5cm 2cm 4cm 0cm}]{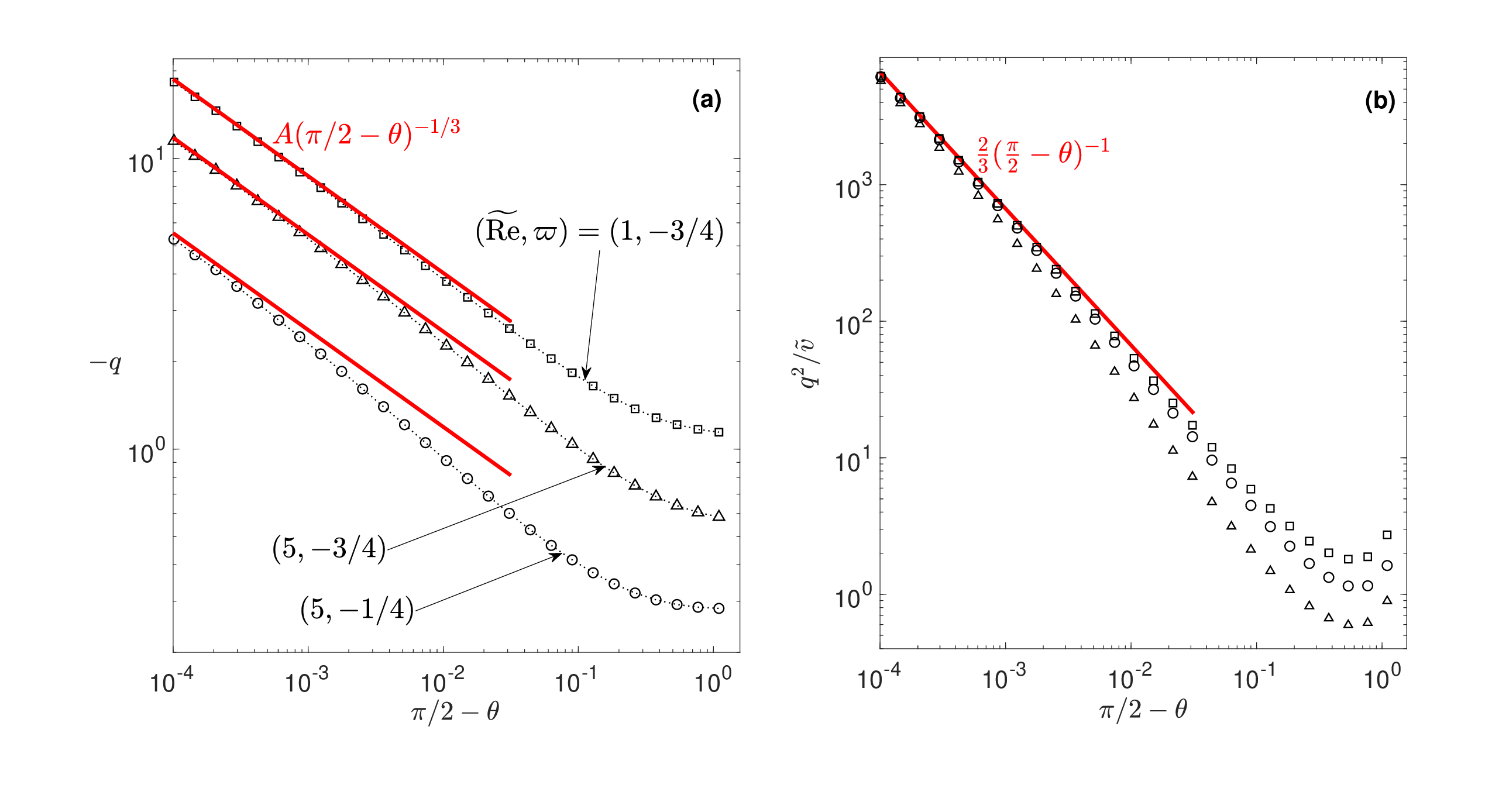}
\caption{(a) Agreement between numerical results for $-q$ and the local behaviour (\ref{blow up}a), for the indicated $(\widetilde{\mathrm{Re}},\varpi)$ parameter pairs; the singularity prefactor $A$ is calculated as part of the numerical solution. (b) Numerical data collapsed on the universal power law \eqref{local collapse} for $q^2/\tilde{v}$.}
\label{fig:singular_local}
\end{center}
\end{figure}
\begin{figure}[t!]
\begin{center}
\includegraphics[scale=0.45,trim={4cm 1cm 4cm 0}]{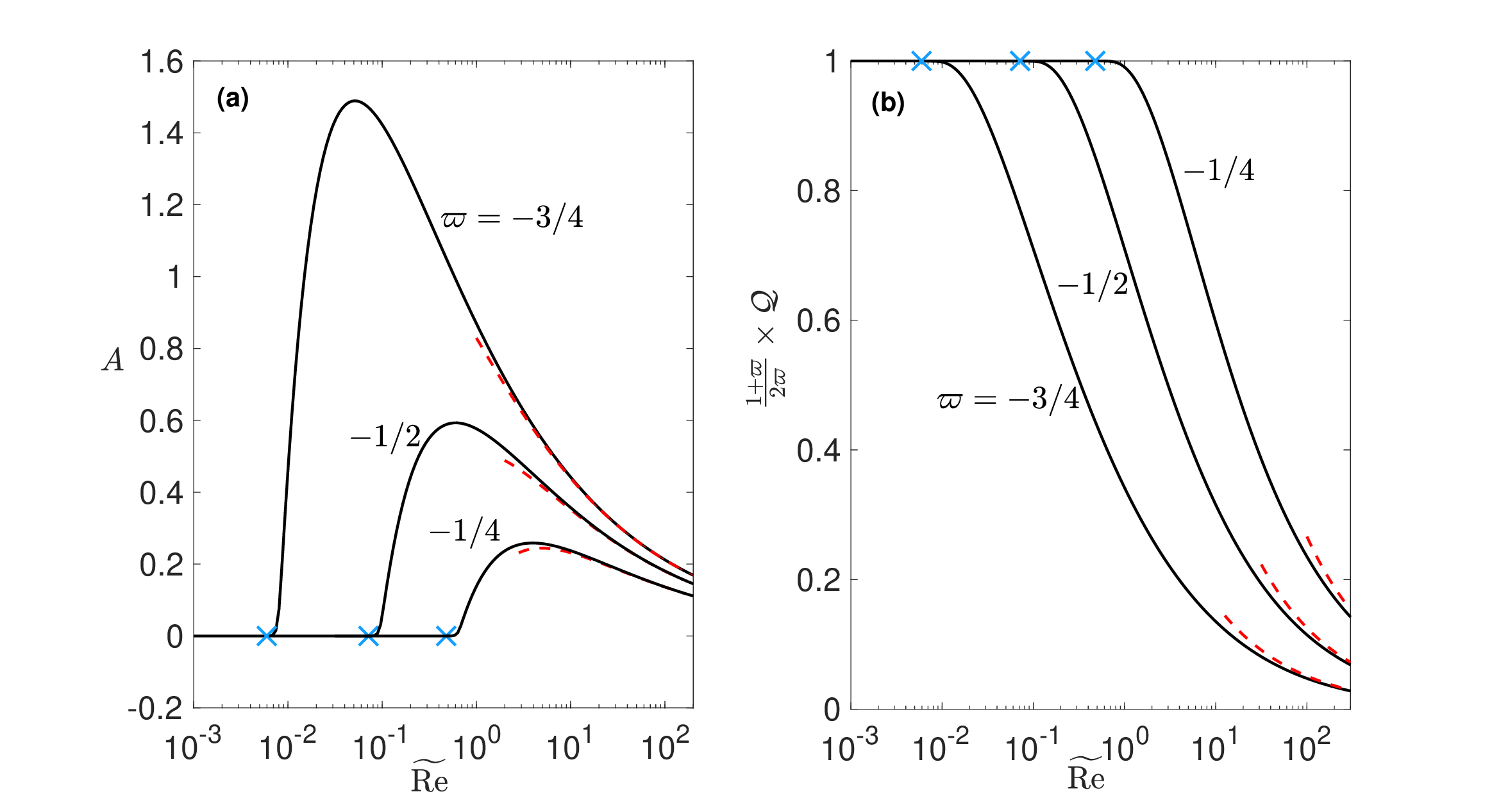}
\caption{(a) Blowup prefactor $A$ [cf.~\eqref{blow up}] and (b) quadrant charge $\mathcal{Q}$, normalized by its pre-blowup value \eqref{q quad regular}, as a function of $\widetilde{\mathrm{Re}}$. Solid curves: numerical solutions. Dashed curves: large-$\widetilde{\mathrm{Re}}$ approximations \eqref{Q asymptotic} and \eqref{A asymptotic}. The crosses mark the onset of the transition regime, which precedes the blowup regime (see Sec.~\ref{ssec:transition})}
\label{fig:bifurcations}
\end{center}
\end{figure}

\subsection{Blowup solutions}
\label{ssec:singularsolutions}
Armed with the local asymptotics developed in Sec.~\ref{sec:local} and global charging--annihilation balance \eqref{charging annal}, in Appendices \ref{app:numerics_singular}-\ref{app:numerics_improved} we develop a modified ``singularity-capturing'' Fourier-series scheme that explicitly allows for the formation of the blowup singularity at the drop equator. This scheme encodes the signatures of the local behaviors \eqref{blow up} in Fourier space, with the singularity prefactor $A$ solved for as part of the problem --- with \eqref{charging annal} serving as the corresponding auxiliary equation. 

Fig.~\ref{fig:omnegative_singular} shows $q$ and $\tilde{v}$ profiles computed using the singularity-capturing scheme described above, in the case $\varpi=-1/4$ and for several values of $\widetilde{\mathrm{Re}}$: $0$, $0.5$, $0.8$ and $100$. This reproduces   Fig.~\ref{fig:omnegative_prelim}, now including valid solutions deep into the blowup regime. The continuous profiles for $\widetilde{\mathrm{Re}}=0$ and $0.5$ are, of course, identical to those already shown in Fig.~\ref{fig:omnegative_prelim}. The profiles for $\widetilde{\mathrm{Re}}=0.8$ and $100$, which exhibit equatorial blowup, are new; note the absence of Gibbs oscillations in the corresponding $q$ profiles, in contrast to the profile for $\widetilde{\mathrm{Re}}=1$ shown in Fig.~\ref{fig:omnegative_prelim}. 
For $\widetilde{\mathrm{Re}}=0.8$ (just past the blowup threshold, as discussed below) $q$ decreases towards the equator excluding a small vicinity of the equator where it diverges. For $\widetilde{\mathrm{Re}}=100$, $q$ and $\tilde{v}$ are much smaller in magnitude (except $q$ very near the equator), with $q$ now diverging monotonically in each quadrant. 

We have checked that the blowup solutions indeed agree with the local structure found in Sec.~\ref{sec:local}. To demonstrate this, we show in 
Fig.~\ref{fig:singular_local}a that the numerically calculated $q$ profiles approach, as $\theta\nearrow\pi/2$, the blowup behaviour (\ref{blow up}a), wherein the singularity prefactor $A$ is determined as part of the  solution. As a further corroboration, Fig.~\ref{fig:singular_local}b demonstrates local collapse of numerical data for the grouping $q^2/\tilde{v}$ upon the universal power-law \eqref{local collapse}.

We next inspect the onset and evolution of the blowup singularity in parameter space.  Fig.~\ref{fig:bifurcations} shows numerical results for the blowup prefactor $A$ and the quadrant charge $\mathcal{Q}$, as a function of $\widetilde{\mathrm{Re}}$ and for several negative values of $\varpi$. We see that $A$ appears to bifurcate from zero at a critical value of $\widetilde{\mathrm{Re}}$, which decreases with increasing $-\varpi$; in accordance with the charging--annihilation balance \eqref{charging annal}, $\mathcal{Q}$ bifurcates at the same value from its pre-blowup value \eqref{q quad regular}. Since our numerical scheme struggles very near the onset of blowup, where $A$ is exceedingly small, we are not able to precisely pinpoint this critical value, which indicates the onset of the blowup regime. For context, we do  mark in Fig.~\ref{fig:bifurcations} the slightly smaller (and precisely computed) critical $\widetilde{\mathrm{Re}}$ values where the solution first becomes non-smooth, i.e., $\alpha$ in \eqref{sublinear} drops below $1$; the transition regime discussed in Sec.~\ref{ssec:transition} corresponds to the narrow $\widetilde{\mathrm{Re}}$ interval bounded by the non-smoothness threshold from below and the blowup threshold from above. 

The $A$ profiles in Fig.~\ref{fig:bifurcations}a feature a maximum as a function of $\widetilde{\mathrm{Re}}$. Nonetheless, the corresponding rate of equatorial annihilation, which is proportional to $\widetilde{\mathrm{Re}}A^3$, increases monotonically with $\widetilde{\mathrm{Re}}$. Indeed, \eqref{charging annal} implies that this corresponds to $\mathcal{Q}$ monotonically decreasing in absolute magnitude, as indeed demonstrated in Fig.~\ref{fig:bifurcations}b. In particular, we see that $\mathcal{Q}$ becomes small for large $\widetilde{\mathrm{Re}}$, in accordance with the small magnitude of $q$ (away from the equator) in the case $(\widetilde{\mathrm{Re}},\varpi)=(100,-1/4)$  presented in Fig.~\ref{fig:omnegative_singular}. 
In the next section, we shall illuminate the regime of large $\widetilde{\mathrm{Re}}$ by means of an asymptotic analysis. 

\section{Blowup at large electric Reynolds numbers}\label{sec:shock_asym}
\subsection{Scaling arguments and leading-order description}\label{ssec:blowup_asym_scalings}
Consider now the limit $\widetilde{\mathrm{Re}}\to\infty$, still assuming the antiparallel polarization scenario $-1<\varpi<0$. In this regime, we expect the solution to exhibit the equatorial blowup singularity analyzed in Sec.~\ref{sec:local}, with $A>0$. We begin by noting that $\mathcal{Q}$ is at most of order unity, i.e., $\mathcal{Q}=\mathcal{O}(1)$. This is because charge annihilation associated with that  blowup singularity can only diminish the absolute magnitude of the quadrant charge $\mathcal{Q}$ from its pre-blowup value \eqref{q quad regular}, which is independent of $\widetilde{\mathrm{Re}}$. The fact that $q$ is strictly negative in the first quadrant [cf.~\eqref{q sign negative}], and therefore does not change its sign, thus implies $q=\mathcal{O}(1)$ --- excluding some small neighbourhood of the equator given its integrable singularity there [cf.~(\ref{blow up}a)]. Thus, the charging condition \eqref{charging in q} suggests, given its right-hand side being independent of $\widetilde{\mathrm{Re}}$, that 
\begin{equation}\label{dominant convection}
\pd{}{\theta}\left(q\tilde{v}\right)=\ord\left(\frac{1}{\widetilde{\mathrm{Re}}}\right).
\end{equation}
In turn, integration of \eqref{dominant convection} using the pole-line symmetry condition (\ref{sym conditions poles}d) suggests    
\begin{equation}\label{dominant convection integrated}
q\tilde{v}=\ord\left(\frac{1}{\widetilde{\mathrm{Re}}}\right).
\end{equation}
Now, since the external field ---  represented in the factorized formulation by the inhomogeneous terms in (\ref{factored equations}a,c) --- is independent of $\widetilde{\mathrm{Re}}$, while $q=\mathcal{O}(1)$, we expect order-unity electric fields. The tangential-stress condition (\ref{factored equations}b) therefore suggests that $\tilde{\mathbf{u}}^{\pm}$ and $q$ have an identical asymptotic scaling, which \eqref{dominant convection integrated} immediately sets as $\widetilde{\mathrm{Re}}^{-1/2}$. Similar scaling arguments were employed by Yariv \& Frankel \cite{Yariv:16}, whose analysis of symmetry-broken Quincke-rotation states in the same two-dimensional setup will be further discussed in Sec.~\ref{sec:discussion}. 

We accordingly posit the expansions
\refstepcounter{equation}
$$
\tilde{\varphi}^{\pm}= \tilde{\varphi}^{\pm}_0 + \widetilde{\mathrm{Re}}^{-1/2}\tilde{\varphi}^{\pm}_{1/2} + \cdots, \quad q=\widetilde{\mathrm{Re}}^{-1/2}q_{1/2} + \cdots, \quad \tilde{\bu}^{\pm}=\widetilde{\mathrm{Re}}^{-1/2}\tilde{\bu}^{\pm}_{1/2} + \cdots.
\label{expansions R<S}
\eqno{(\theequation\mathrm{a},\mathrm{b},\mathrm{c})}
$$ 
With $q$ small, the condition (\ref{factored equations}{a}) reduces at leading order to
\begin{equation}
\pd{\tilde{\varphi}_0^-}{r}=-2\cos\theta \quad \text{at} \quad r=1. 
\end{equation}
The Taylor-symmetric solution of Laplace's equation inside the unit circle that satisfies that condition is
\begin{equation}\label{uni inside}
\tilde{\varphi}_0^-=-2r\cos\theta,
\end{equation}
representing a uniform field. The corresponding exterior potential $\tilde{\varphi}_0^+$ can be obtained from the reflection relation \eqref{pot reflection} as $-2r^{-1}\cos\theta$. 

To find $q_{1/2}$, consider the charging condition \eqref{charging in q} at order unity,
\begin{equation}
\pd{}{\theta}\left(\tilde{v}_{1/2}q_{1/2}\right)=2\varpi\cos\theta\quad \text{at} \quad r=1.
\end{equation}
Due to the symmetry about the pole line, $\tilde{v}_{1/2}$ vanishes at $\theta=0$ [cf.~(\ref{sym conditions poles}d)]. 
Integration from that angle thus gives
\begin{equation}\label{product half half}
q_{1/2}\tilde{v}_{1/2}=2\varpi\sin\theta\quad \text{at} \quad r=1,
\end{equation}
which, upon substitution into the tangential-stress condition (\ref{factored equations}{b}),  yields 
\begin{equation}
\tilde{v}_{1/2}\left(\pd{\tilde{v}_{1/2}^-}{r}-\tilde{v}_{1/2}\right)=-4\varpi\sin^2\theta
\quad \text{at} \quad r=1.
\label{nonlinear shear}
\end{equation}

The problem for the velocity field $\tilde{\bu}_{1/2}^-$ consists of the Stokes equations in the unit disk, regularity at  $r=0$, impermeability at $r=1$, and the nonlinear boundary condition \eqref{nonlinear shear}. Once the solution is obtained, the exterior velocity field $\tilde{\mathbf{u}}^+$ follows from the reflection relation \eqref{psi reflection}, while the surface-charge density $q_{1/2}$ can be deduced from \eqref{product half half}. It is readily seen that this leading-order flow problem is invariant under reversal of the velocity field. This indeterminacy is  resolved, however, by noting that the polarity result \eqref{q sign negative} in conjunction with expression \eqref{product half half} for the surface current  imply that for $\varpi<0$ the correct flow solution must satisfy 
\begin{equation}
\tilde{v}_{1/2}>0 \quad \text{for} \quad 0<\theta<\pi/2, \label{assert}
\end{equation}
consistently with the schematic picture in Fig.~\ref{fig:scenarios}. 

The above deduction raises the question whether the same asymptotic structure may also apply in the parallel-polarization case $\varpi>0$, but with $\tilde{v}_{1/2}<0$ for $0<\theta<\pi/2$. To rule out this possibility, we note (following Yariv \& Frankel \cite{Yariv:16}) that the integral of $\tilde{v}_{1/2}(\partial\tilde{v}_{1/2}^-/\partial{r}-\tilde{v}_{1/2})$ over the drop interface, representing mechanical power, must be non-negative, so that viscous dissipation within the drop is non-negative. Accordingly, the nonlinear boundary condition \eqref{nonlinear shear} yields the condition $\varpi<0$ for a non-trivial solution. An alternative asymptotic scheme appropriate to the parallel-polarization scenario will be developed in Sec.~\ref{sec:caps}.  

\subsection{Existence of an inner region near the equator}\label{ssec:inner_existence}
Consider the local behaviors of the asymptotic fields as the distance from the equator $\rho\searrow0$. We see from \eqref{product half half} that the product ${q}_{1/2}\tilde{v}_{1/2}$, representing a rescaled surface current, approaches a finite non-vanishing value in that limit. Hence, the nonlinear boundary condition \eqref{nonlinear shear} implies that ${q}_{1/2}$ and ${v}_{1/2}$ scale as $\rho^{-1/2}$ and $\rho^{1/2}$, respectively. (A detailed local analysis of the asymptotic fields will be carried out in subsection \ref{ssec:universalflow}.) 

Clearly, this local structure of the asymptotic fields differs from that found in Sec.~\ref{sec:local} based on local analysis of the exact problem formulation for \emph{fixed} $\widetilde{\mathrm{Re}}$; while both describe blowup with surface current approaching a finite limit, the blowup is less singular in the fixed-$\widetilde{\mathrm{Re}}$ local structure, with the charge density scaling as $\rho^{-1/3}$. This suggests that the large-$\widetilde{\mathrm{Re}}$ expansions \eqref{expansions R<S} are not uniformly valid --- in particular, that they constitute ``outer'' (drop-scale) expansions that asymptotically match with an ``inner'' region near the equator \cite{Hinch:book}. The need for an inner region can also be inferred directly. Indeed, Gauss's law (\ref{factored equations}a) and the form of the asymptotic expansions \eqref{expansions R<S} suggest that the electric field corresponding to $\tilde{\varphi}^-_{1/2}$ scales similarly to $q_{1/2}$, i.e., as $\rho^{-1/2}$. That locally induced field therefore becomes comparable to the global uniform field \eqref{uni inside} for $\rho=\mathcal{O}(1/\widetilde{\mathrm{Re}})$, indicating an inner region of that width. We note that asymptotic matching implies that $q$ is of order unity there.

\subsection{Universal flow problem}
\label{ssec:universalflow}
The problem governing $\tilde{\bu}_{1/2}^-$ depends upon the single parameter $\varpi$. That dependence may be eliminated by defining
\begin{equation}
\tilde{\bu}_{1/2}^-=(-\varpi)^{1/2}\acute{\bu}^{-}, 
\end{equation} 
and similarly for the corresponding exterior velocity field $\tilde{\bu}_{1/2}^+$, whereby the nonlinear boundary condition  \eqref{nonlinear shear} becomes  
\begin{equation}
\acute{v}\left(\pd{\acute{v}^-}{r}-\acute{v}\right)=4\sin^2\theta\quad \text{at} \quad r=1.
\label{universal BC}
\end{equation}
Thus, the flow field $\acute{\bu}^-$ satisfies a \emph{universal} flow problem, consisting 
of the Stokes equations within the unit circle, regularity at $r=0$, the impermeability condition,
\begin{equation}
\acute{u}^- = 0 \quad \text{at} \quad r=1,
\end{equation}
and the nonlinear boundary condition \eqref{universal BC}. Furthermore, writing
\begin{equation}
q_{1/2}=(-\varpi)^{1/2}\acute{q},
\end{equation}
we find from \eqref{product half half} the  corresponding universal surface-charge density  
\begin{equation}
\acute{q}=-2\frac{\sin\theta}{\acute{v}}. \label{universal charge density}
\end{equation}

\begin{figure}[t!]
\begin{center}
\includegraphics[scale=0.45,trim={4cm 1cm 4cm 0}]{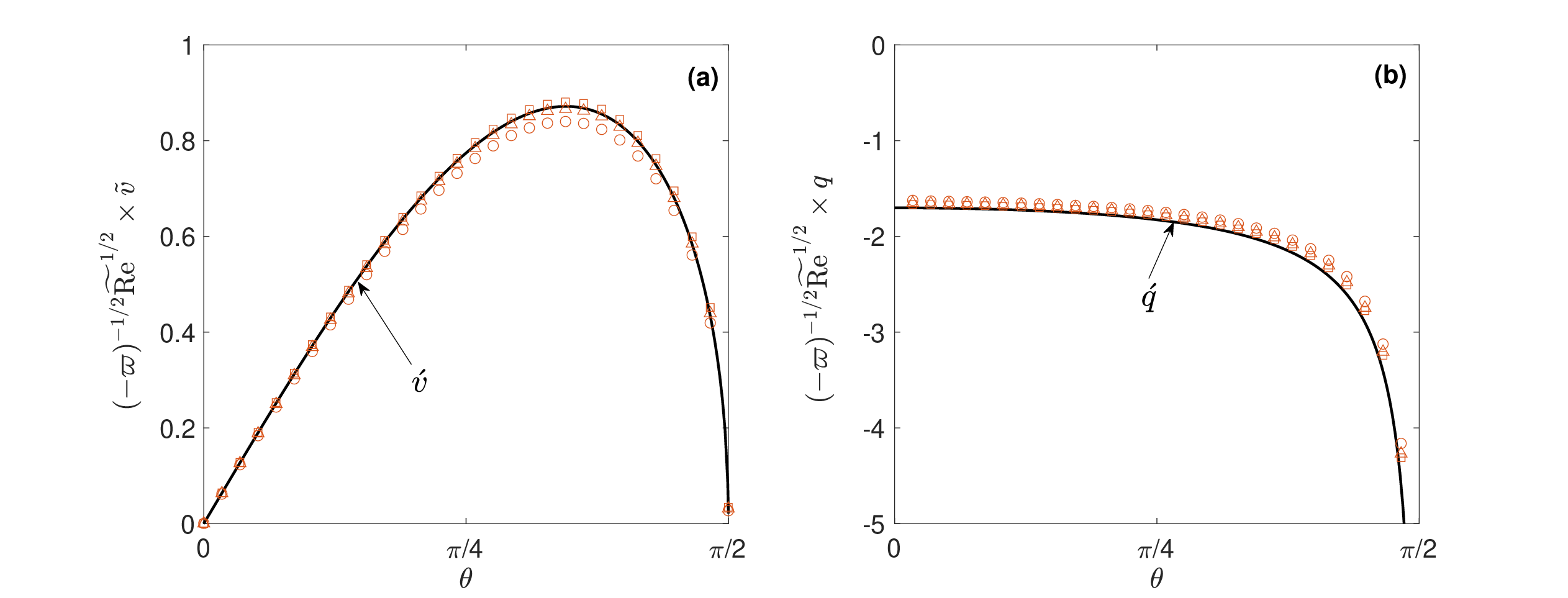}
\caption{Scaled (a) surface velocity and (b) charge density. Solid curves: Profiles $\acute{v}$ and $\acute{q}$ obtained from numerical solution of the large-$\widetilde{\mathrm{Re}}$ universal problem (Sec.~\ref{sec:shock_asym}). Symbols: full numerics at $\widetilde{\mathrm{Re}}=300$ for $\varpi=-1/4$ (circles), $\varpi=-1/2$ (triangles) and $\varpi=-3/4$ (squares).}
\label{fig:acute}
\end{center}
\end{figure}

Since the above problem for $\tilde{\bu}^-$ is parameter-free, it only needs to be solved once. We used a straightforward collocation method wherein the stream function corresponding to $\acute{\bu}^-$ is represented by an appropriate Fourier-series solution of the biharmonic equation [cf.~\eqref{psi fourier}]. In Fig.~\ref{fig:acute}, we present the surface velocity $\acute{v}$ obtained from that solution and the corresponding surface-charge density \eqref{universal charge density}. The universality of these profiles is demonstrated by comparison with full numerics at $\widetilde{\mathrm{Re}}=300$, for several negative values of $\varpi$.  

We next derive the leading-order behavior near $\theta=\pi/2$ of the stream function $\acute{\psi}^-$ associated with $\acute{\bu}^-$. To this end, we employ the local polar coordinates $(\rho,\phi)$ defined in Sec.~\ref{sec:local}, with $\rho\searrow0$. Given the scalings determined in Sec.~\ref{ssec:inner_existence}, we seek a biharmonic function that is proportional to $\rho^{3/2}$, odd in $\phi$, and satisfies interfacial impermeability, $\acute\psi^- = 0$ at $\phi=\pi/2$, and the nonlinear boundary condition \eqref{universal BC} which assumes the form
\begin{equation}
\pd{\acute\psi^-}{\phi}\pd{^2\acute\psi^-}{\phi^2} = 4\rho^3
\quad \text{at} \quad \phi= \pi/2. \label{nonlinear shear acute psi}
\end{equation} 
The solution, 
\begin{equation}
\acute\psi^- = 2^{1/2}\rho^{3/2} \left(\sin\frac{3\phi}{2}-\sin\frac{\phi}{2}\right),
\end{equation}
implies the local behavior
\begin{equation}
\acute v \sim 2(\pi/2-\theta)^{1/2}\quad \text{as} \quad \theta\nearrow\pi/2.
\label{blow up universal v}
\end{equation}
We then find from \eqref{universal charge density} 
\begin{equation}
\acute q \sim -(\pi/2-\theta)^{-1/2}\quad \text{as} \quad \theta\nearrow\pi/2.
\label{blow up universal q}
\end{equation}

\begin{figure}[t!]
\begin{center}
\includegraphics[scale=0.43,trim={4cm 1cm 4cm 0}]{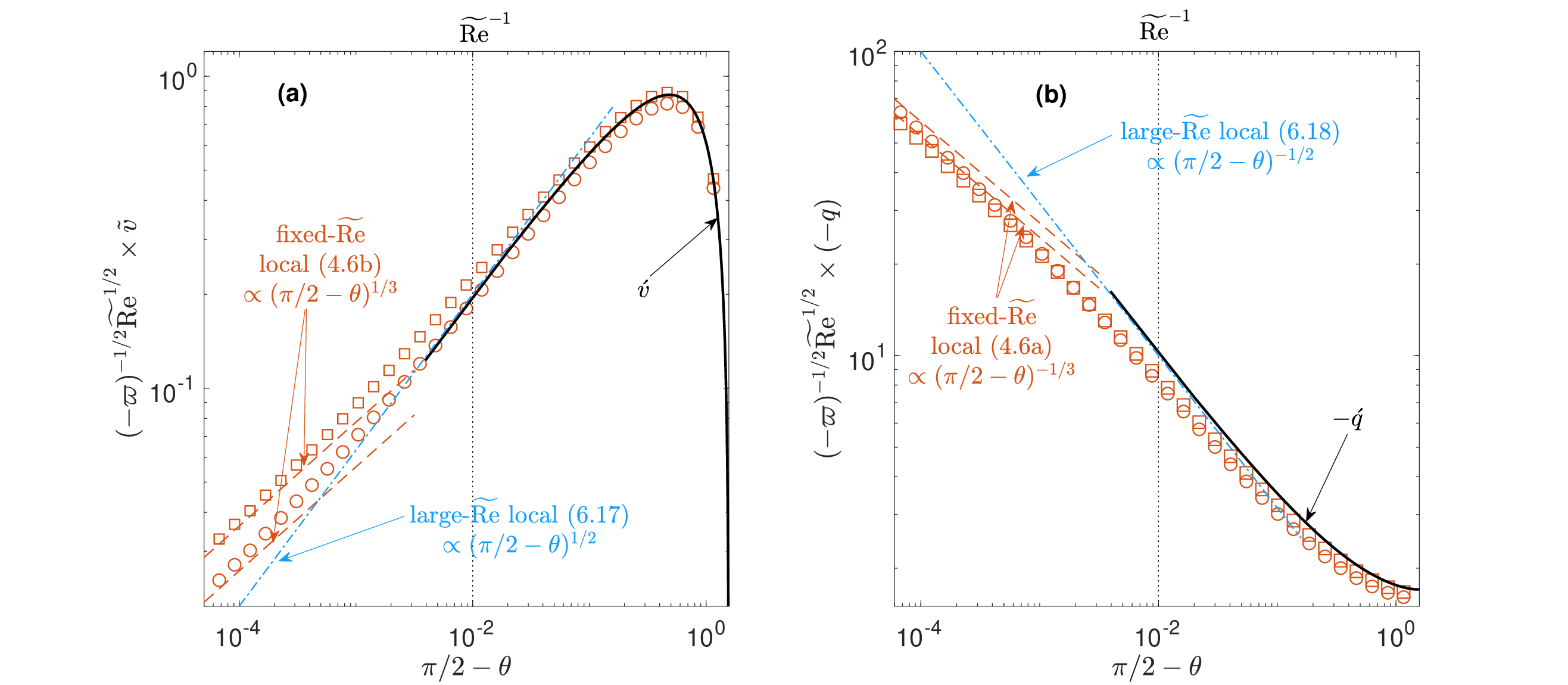}
\caption{Scaled (a) surface velocity and (b) charge density. Solid curves: Profiles $\acute{v}$ and $\acute{q}$ obtained from numerical solution of the large-$\widetilde{\mathrm{Re}}$ universal problem (Sec.~\ref{sec:shock_asym}). Symbols: full numerics at $\widetilde{\mathrm{Re}}=100$ for $\varpi=-1/4$ (circles) and $\varpi=-3/4$ (squares). Dash-dotted curves: local approximations of the universal profiles [see \eqref{blow up universal v}--\eqref{blow up universal q}]. Dashed curves: local approximations at fixed $\widetilde{\mathrm{Re}}$ [see \eqref{blow up}].
}
\label{fig:localtransition}
\end{center}
\end{figure}

As we have anticipated in subsection \ref{ssec:inner_existence}, the local behaviors \eqref{blow up universal v} and \eqref{blow up universal q} of the large-$\widetilde{\mathrm{Re}}$ profiles do not agree with the fixed-$\widetilde{\mathrm{Re}}$ local approximations \eqref{blow up} derived in Sec.~\ref{sec:local}. In Fig.~\ref{fig:localtransition}, we illustrate the transition between the two local behaviours across the $\mathcal{O}(\widetilde{\mathrm{Re}}^{-1})$ inner region about the equator identified in that subsection. We note that the collapse of the data for different $\varpi$ on the universal large-$\widetilde{\mathrm{Re}}$ profiles, which is especially evident in Fig.~\ref{fig:acute}, breaks down near the equator. 

\begin{figure}[t!]
\begin{center}
\includegraphics[scale=0.4,trim={1cm 1cm 0 0}]{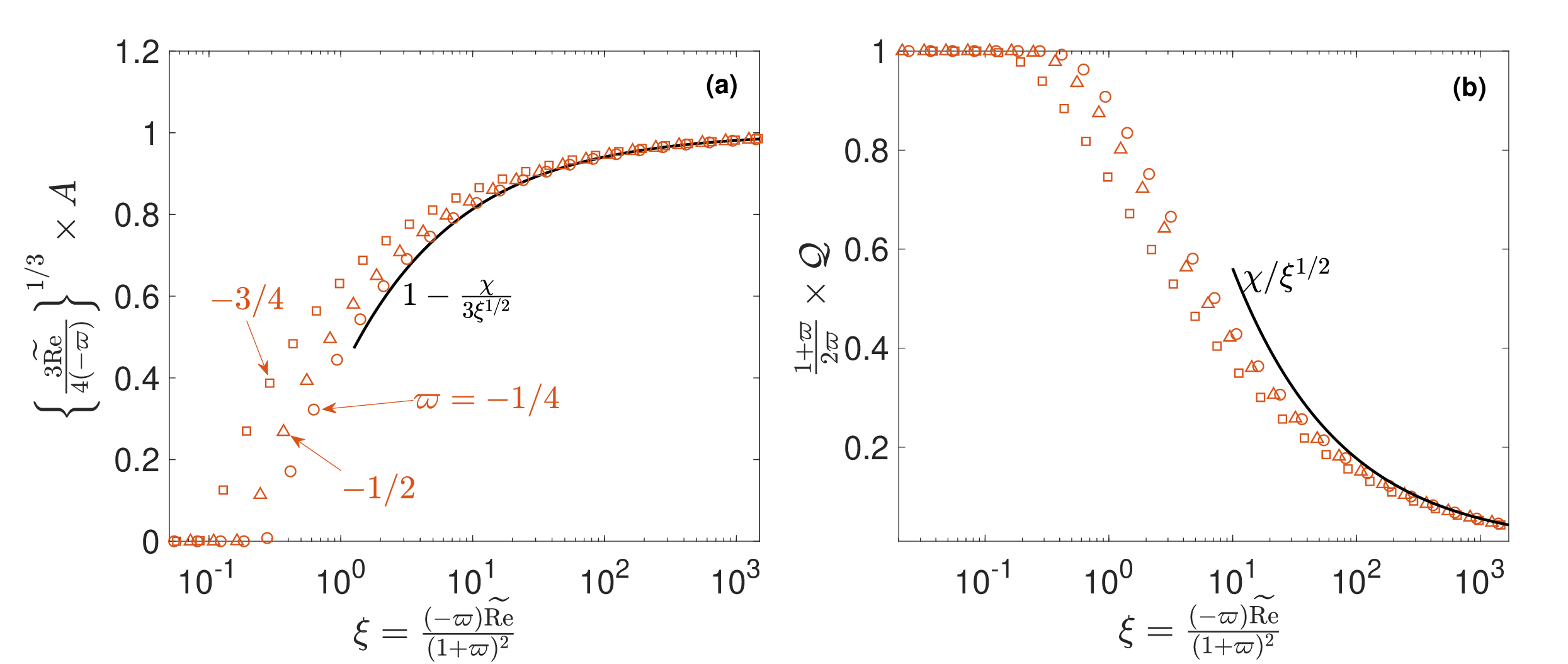}
\caption{(a) Singularity prefactor $A$, scaled by $\{(-4\varpi)/(3\widetilde{\mathrm{Re}})\}^{1/3}$, and (b) quadrant charge $\mathcal{Q}$, scaled by the pre-blowup value $2\varpi/(1+\varpi)$ [cf.~\eqref{q quad regular}], as a function of the grouping $\xi=(-\varpi\widetilde{\mathrm{Re}})/(1+\varpi)^2$. Solid lines: $\widetilde{\mathrm{Re}}\gg1$ asymptotic approximations \eqref{Q asymptotic} and \eqref{A asymptotic}. Symbols: numerical results for $\varpi=-1/4$ (circles), $\varpi=-1/2$ (triangles) and $\varpi=-3/4$ (squares).}
\label{fig:AQcollapse}
\end{center}
\end{figure}
\subsection{Quadrant charge and blowup prefactor} 
In accordance with the discussion in  subsection \ref{ssec:inner_existence}, the universal solution constitutes a leading-order outer approximation, which breaks down in an inner region near the equator. Accordingly, the asymptotic behaviors \eqref{blow up universal v} and \eqref{blow up universal q} represent local approximations of the outer solution, whereas the true local structure \eqref{blow up} is hidden in the inner region.
Remarkably, the prefactor $A$ determining that blowup structure can nonetheless be extracted directly from the outer universal solution, thus circumventing the need for a detailed inner analysis. Indeed, given the inverse-square-root divergence \eqref{blow up universal q}, the universal charge density is integrable over $(0,\pi/2)$, implying a net charge of order $\widetilde{\mathrm{Re}}^{-1/2}$. Being asymptotically larger than the net charge associated with the inner region [accumulated by an $\ord(1)$ density over an $\mathcal{O}(1/\widetilde{\mathrm{Re}})$ angle], it provides a leading-order approximation for the quadrant charge $\mathcal{Q}$. Recalling \eqref{universal charge density}, we therefore obtain
\begin{equation}\label{Q asymptotic}
\mathcal{Q} \sim -2\chi(-\varpi)^{1/2}   \widetilde{\mathrm{Re}}^{-1/2}   \quad \text{as} \quad \widetilde{\mathrm{Re}}\to\infty,
\end{equation}
wherein the pure number
\begin{equation}
\chi=\int_0^{\pi/2}\frac{\sin\theta}{\acute{v}} \, \dd\theta
\end{equation} 
may be obtained from the numerical solution of the universal problem as $\approx 1.775\ldots$. Plugging into
\eqref{A in Q} yields the two-term approximation
\begin{equation}\label{A asymptotic}
A \sim \left(\frac{-4\varpi}{3\widetilde{\mathrm{Re}}}\right)^{1/3}
\left[1-\frac{(1+\varpi)\chi}{3(-\varpi)^{1/2}\widetilde{\mathrm{Re}}^{1/2}} + \cdots \right]
 \quad \text{as} \quad \widetilde{\mathrm{Re}}\to\infty.
\end{equation}

The approximations \eqref{Q asymptotic} and \eqref{A asymptotic} are depicted in Fig.~\ref{fig:bifurcations}, which as discussed in Sec.~\ref{ssec:singularsolutions} shows numerical data for $A$ and $\mathcal{Q}$ for several negative values of $\varpi$. In Fig.~\ref{fig:AQcollapse}, we show how that data can be collapsed at large $\widetilde{\mathrm{Re}}$ using the asymptotic approximations. Indeed, upon defining the grouping $\xi=(-\varpi\widetilde{\mathrm{Re}})/(1+\varpi)^2$, we find from \eqref{Q asymptotic} and \eqref{A asymptotic} that $A$, scaled by $\{(-4\varpi)/(3\widetilde{\mathrm{Re}})\}^{1/3}$, and $\mathcal{Q}$, scaled by $2\varpi/(1+\varpi)$, possess the universal asymptotic expansions $\sim 1 - \chi/(3\xi^{1/2})$ and $\sim \chi/\xi^{1/2}$ as $\xi\to\infty$, respectively. In recasting \eqref{Q asymptotic} as a universal approximation, there is freedom in how to rescale $\mathcal{Q}$. The present choice ensures collapse also for pre-blowup $\widetilde{\mathrm{Re}}$ --- recall \eqref{q quad regular}. Fortuitously, it also leads to the same $\xi$-rescaling of $\widetilde{\mathrm{Re}}$ required for recasting \eqref{A asymptotic} as a universal expansion for $A$.  

\section{Polar caps at large electric Reynolds numbers}
\label{sec:caps}
In this section, we consider the case $\varpi>0$ where the charge density is polarized parallel to the external field [cf.~\eqref{q sign positive}]. In contrast to the antiparallel-polarization scenario corresponding to the case $-1<\varpi<0$, we do not observe any form of singularity in the numerical simulations; in particular, we do not encounter the blowup singularity studied in Sec.~\ref{sec:local}. Accordingly, charge annihilation does not occur.  This suggests a focus on the asymptotic regime $\widetilde{\mathrm{Re}}\gg1$. 
\begin{figure}[t!]
\begin{center}
\includegraphics[scale=0.38,trim={5cm 1.7cm 4cm 0}]{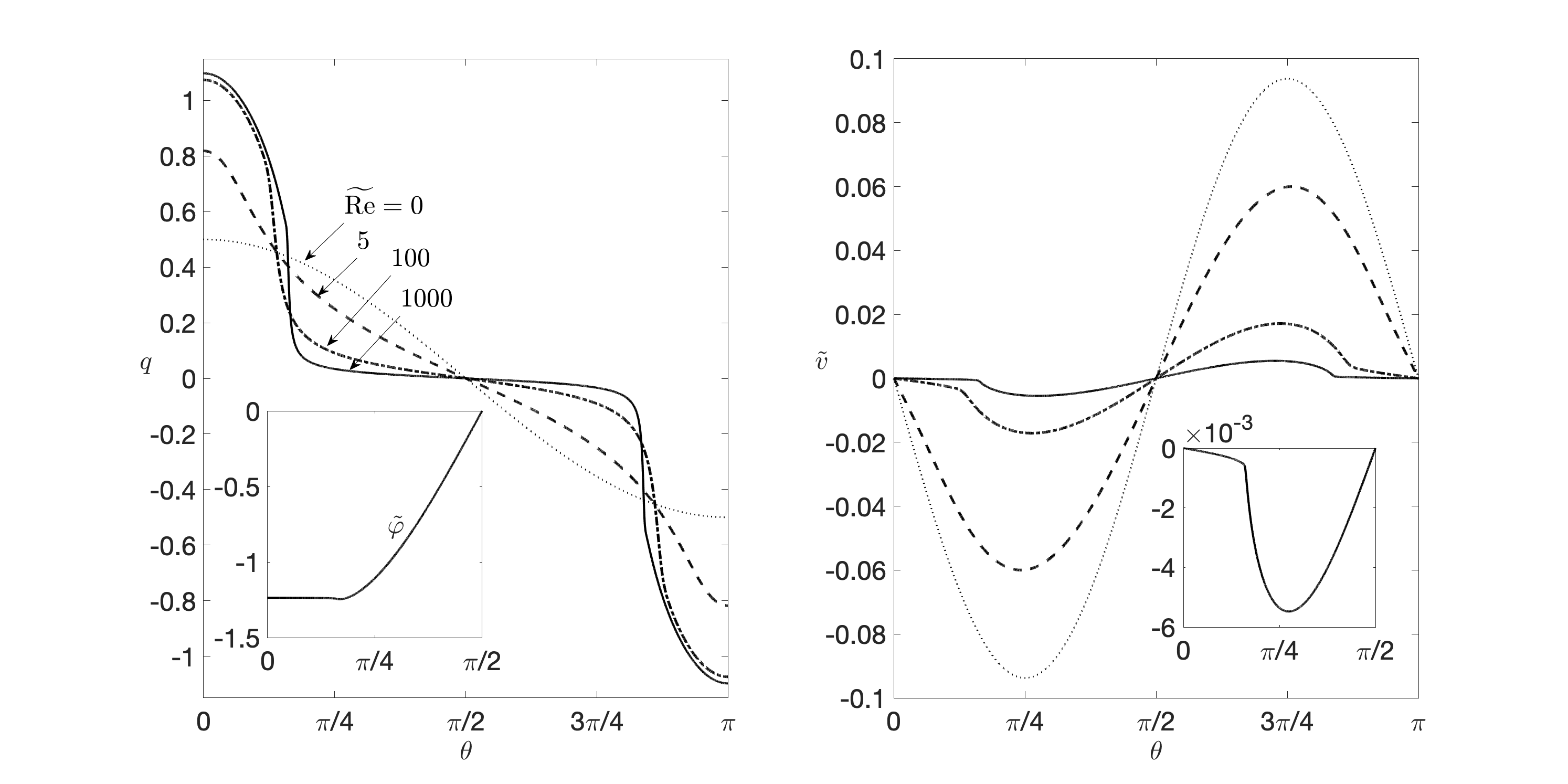}
\caption{(a) Charge density $q$ and (b) surface velocity $\tilde{v}$  for $\varpi=1/3$ and several values of $\widetilde{\mathrm{Re}}$: 0 (dotted), 5 (dashed), 100 (dash-dotted) and 1000 (solid); for the latter value, the inset of (a) shows surface potential while that of (b) depicts the azimuthal velocity more clearly.} 
\label{fig:ompositive}
\end{center}
\end{figure}
\subsection{Numerical observations}
\label{ssec:cap_numerics}
\begin{figure}[t!]
\begin{center}
\includegraphics[scale=0.45,trim={4cm 2cm 4cm 0}]{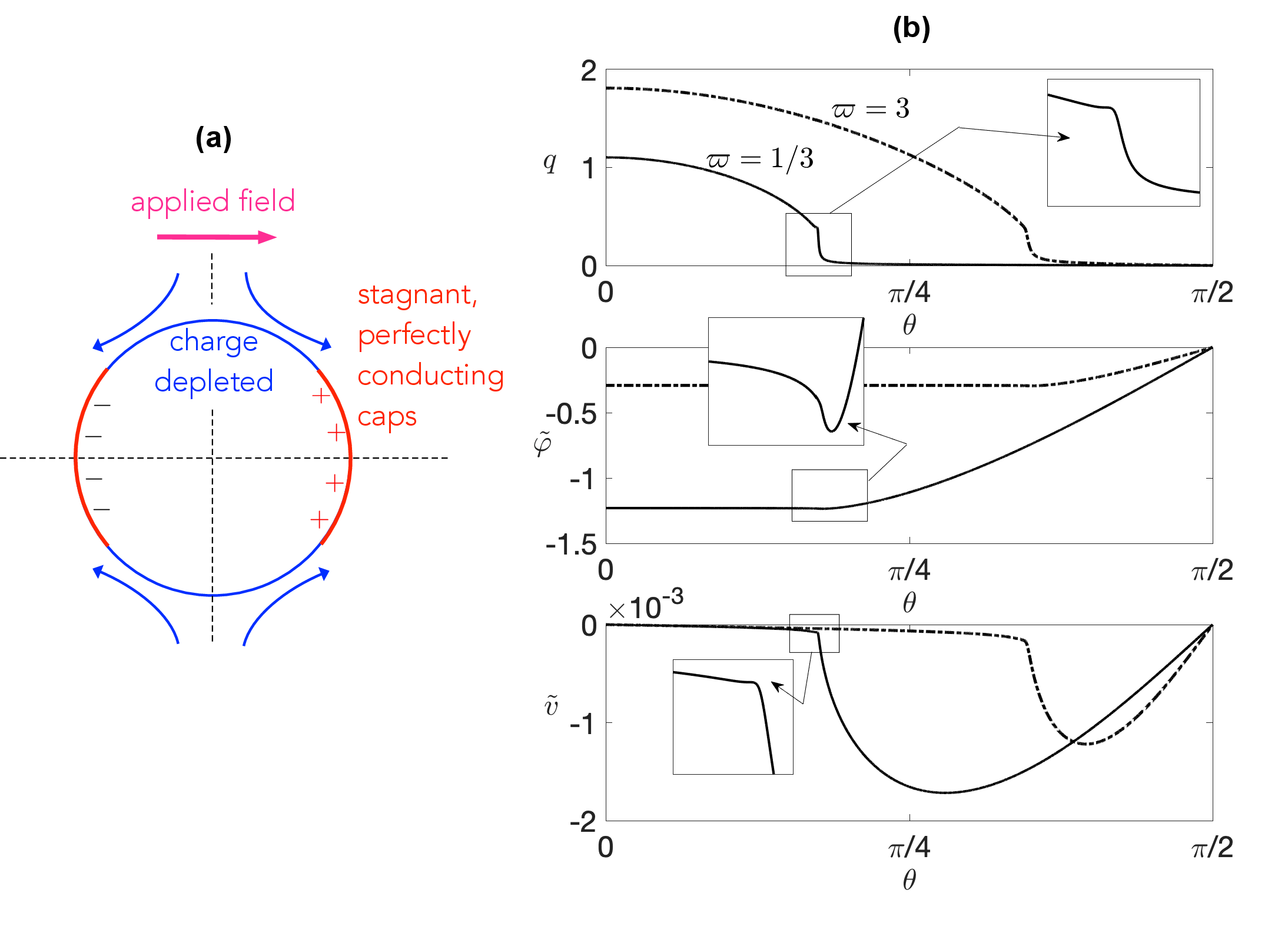}
\caption{(a) Schematic of the stagnant, perfectly conducting caps observed in the parallel-polarization case $\varpi>0$ at large $\widetilde{\mathrm{Re}}$. (b) Charge density $q$, surface potential $\tilde{\varphi}$ and surface velocity $\tilde{v}$ for $\widetilde{\mathrm{Re}}=10000$ and two values of $\varpi$: 1/3 (solid) and 3 (dash-dotted); the insets are included in order to demonstrate the smoothness of the profiles.}
\label{fig:ompositive_largeRe}
\end{center}
\end{figure}
We begin by presenting and discussing preliminary numerical observations (obtained using the straightforward Fourier-series scheme described in Appendix \ref{app:numerics}). In Fig.~\ref{fig:ompositive}, we show the charge density $q$ and surface velocity $\tilde{v}$  for $\varpi=1/3$ and several values of $\widetilde{\mathrm{Re}}$: $0$, $5$, $100$ and $1000$. We note that the sign of $q$ over the front and back halves of the drop are as predicted by the polarity result \eqref{q sign positive}, and that the direction of the surface velocity is from the equator to the poles in all of the cases shown. With increasing $\widetilde{\mathrm{Re}}$, $q$ becomes increasingly concentrated about the drop poles at $\theta=0,\pi$, while $\tilde{v}$ generally decreases in magnitude. At large $\widetilde{\mathrm{Re}}$, $q$ is essentially confined to regions centered about the poles wherein the surface potential $\tilde{\varphi}$ is approximately uniform and $\tilde{v}$ is relatively small (see insets in Fig.~\ref{fig:ompositive}). As depicted schematically in Fig.~\ref{fig:ompositive_largeRe}a, we interpret these regions as surface-charge ``caps'' which are hydrodynamically stagnant and electrically perfectly conducting; this picture is reinforced by the numerical results presented in Fig.~\ref{fig:ompositive_largeRe}b for $\widetilde{\mathrm{Re}}=10^4$ and two values of $\varpi$: $1/3$ and $3$. Comparison of the results for those two values suggests that cap size increases with $\varpi$. The insets, zooming in on the cap edges, demonstrate that the solutions remain smooth despite their non-smooth appearance on the drop scale. 

\subsection{Asymptotic structure}
\label{ssec:cap_structure}
The remainder of this section is devoted to an asymptotic analysis of the parallel-polarization case $\varpi>0$ in the limit $\widetilde{\mathrm{Re}}\to\infty$, with the aim of illuminating the formation of surface-charge caps as we have numerically observed. In the absence of charge annihilation, the quadrant charge $\mathcal{Q}$ is explicitly provided by \eqref{q quad regular}, a function of $\varpi$ alone. For convenience, we restrict the analysis in this section to the first quadrant $\theta\in(0,\pi/2)$, with symmetry conditions applied at $\theta=0$ and $\pi/2$ [cf.~\eqref{sym conditions poles} and \eqref{sym conditions equator}]. 

In Sec.~\ref{sec:shock_asym}, we have argued based on a dissipation argument that the asymptotic scheme developed in that section is inconsistent with parallel polarization. This is also evident by the impossibility of an asymptotically small $q$ scaling in the absence of charge annihilation. Nonetheless, the scaling arguments leading to that scheme do not rely on the polarization scenario. In particular, the asymptotic estimate $q\tilde{v}=\ord(1/\widetilde{\mathrm{Re}})$ [cf.~\eqref{dominant convection integrated}] stands up to scrutiny --- in fact, its justification is simplified by the absence of charge annihilation.

With $\mathcal{Q}$ given by \eqref{q quad regular} and $q$ being strictly positive in the first quadrant $\theta\in(0,\pi/2)$ [cf.~\eqref{q sign positive}], one might naively assume $q=\text{ord}(1)$ holds for that domain. Assuming so, the estimate \eqref{dominant convection integrated} gives $\tilde{v}=\ord(1/\protect \widetilde{\mathrm{Re}})$. The shear-stress condition (\ref{factored equations}b), in conjunction with the symmetry condition (\ref{sym conditions equator}b), then gives $\tilde{\varphi}=o(1)$. Given (\ref{factored equations}a), however, the latter result implies $q\sim 2\cos\theta$, which contradicts \eqref{q quad regular}. 

Thus, on one hand, $q$ cannot be $o(1)$ as argued in Sec.~\ref{sec:shock_asym}, while on the other hand the assumption $q=\text{ord}(1)$ results in a contradiction. The resolution of this apparent paradox is evident from our numerical observations --- parts of the interface are appreciably charged,  $q=\text{ord}(1)$, whereas the remaining parts are charge-depleted, $q=o(1)$. (In retrospect, then, the problem with both scaling arguments is that they tacitly assume uniform scalings of the surface fields.) In accordance with those observations, the simplest such decomposition of the interface which is compatible with the Taylor symmetry has the charged part of the interface consisting of a pair of circular-arc intervals (``caps'') centered about the drop poles. We denote the opening angle of these caps by $2\theta^*$. As shown in Fig.~\ref{fig:cap_asymptotics}a, we shall denote by $\mathcal{C}$ the subset $\theta\in (0,\theta^*)$ of the first quadrant coinciding with the front cap, and by $\overline{\mathcal{C}}$ the complementary subset $\theta\in (\theta^*,\pi/2)$. 

What is the scaling of $q$ in $\overline{\mathcal{C}}$? Denoting it by $\lambda\ll1$, \eqref{dominant convection integrated} suggests that $\tilde{v}$ is order $(\lambda\widetilde{\mathrm{Re}})^{-1}$ on $\overline{\mathcal{C}}$ --- larger than the order $1/\widetilde{\mathrm{Re}}$ velocity on ${\mathcal{C}}$. Accordingly, the velocity fields $\tilde{\bu}^{\pm}$ are of order $(\lambda\widetilde{\mathrm{Re}})^{-1}$ and, assuming that these fields vary on the drop scale, the shear stresses $\partial{\tilde{v}^{\pm}}/\partial r-\tilde{v}$ possess the same scaling. It follows that the leading-order velocity fields satisfy a no-slip condition on $\mathcal{C}$ --- the surface-charge caps are hydrodynamically ``stagnant.''  
Given the scalings for $q$, consideration of the problem governing the interior potential $\tilde{\varphi}^-$ (cf.~Sec.~\ref{ssec:reduced}) suggests it is of order unity (and hence so is $\tilde{\varphi}^+$). Thus, assuming variations of the potential on the drop scale, the tangential-stress balance (\ref{factored equations}b) on $\overline{\mathcal{C}}$ yields $\lambda=\widetilde{\mathrm{Re}}^{-1/2}$. 

Accordingly, we posit the asymptotic expansions
\refstepcounter{equation}
$$
\label{cap electrical expansions}
\tilde{\varphi}^{\pm}= \tilde{\varphi}^{\pm}_0 + \widetilde{\mathrm{Re}}^{-1/2}\tilde{\varphi}^{\pm}_{1/2} + \cdots, \quad q=q_0+\widetilde{\mathrm{Re}}^{-1/2}q_{1/2} + \cdots,
\eqno{(\theequation \mathrm{a},\mathrm{b})}
$$
with $q_0$ vanishing on $\overline{\mathcal{C}}$, 
and
\begin{equation}
\label{cap flow expansions}
\tilde{\bu}^{\pm}=\widetilde{\mathrm{Re}}^{-1/2}\tilde{\bu}^{\pm}_{1/2} + \widetilde{\mathrm{Re}}^{-1}\tilde{\bu}^{\pm}_1 + \cdots,
\end{equation}
 with $\tilde{\bu}_{1/2}^{\pm}$ vanishing on $\mathcal{C}$.  Furthermore, consideration of the tangential-stress balance (\ref{factored equations}b) on $\mathcal{C}$, noting that the tangential stress is of order $\widetilde{\mathrm{Re}}^{-1/2}$ and $q$ is of order unity on $\mathcal{C}$, gives that $\partial{\tilde{\varphi}}/\partial\theta$ is small on $\mathcal{C}$, of order $\widetilde{\mathrm{Re}}^{-1/2}$. It follows that $\partial{\tilde{\varphi}_0}/\partial\theta=0$ on $\mathcal{C}$, namely the surface-charge caps are to leading order perfectly conducting. 

Despite the smooth appearance of the numerical solutions in the case $\varpi>0$, even for very large $\widetilde{\mathrm{Re}}$ (see Fig.~\ref{fig:ompositive_largeRe}), we expect the leading-order fields $\tilde{\varphi}^{\pm}_0$, $q_0$ and $\tilde{\bu}^{\pm}_{1/2}$ to be piecewise smooth over the interface given that they are to satisfy mixed boundary conditions there. We accordingly interpret \eqref{cap electrical expansions} and \eqref{cap flow expansions} as ``outer'' asymptotic expansions  and conjecture that the smoothness of the solutions could be demonstrated by asymptotic matching with ``inner'' regions near the cap edges. We shall further discuss these inner regions in subsection \ref{ssec:cap_hydrodynamic}.

We next go beyond scaling arguments to formulate a pair of mixed boundary-value  problems governing the leading-order electric and hydrodynamic fields, respectively. As we shall see, these problems can be formulated and solved sequentially. 

\subsection{Electric cap problem}
\label{ssec:cap_electric}
We begin by formulating a problem governing the leading-order charge density $q_0$ and interior potential $\tilde{\varphi}^{-}_0$. [The corresponding exterior potential $\tilde{\varphi}^+_0$ can be obtained from $\tilde{\varphi}^-_0$ using the reflection relation \eqref{pot reflection}.] These fields are related according to the leading-order balance of the factorized Gauss's law (\ref{factored equations}a), 
\begin{equation}\label{q0 and phi0}
q_0=\pd{\tilde{\varphi}_0^-}{r}+2\cos\theta \quad \text{at} \quad r=1.
\end{equation}

\begin{figure}[t!]
\begin{center}
\includegraphics[scale=0.45,trim={4cm 2cm 2cm 0}]{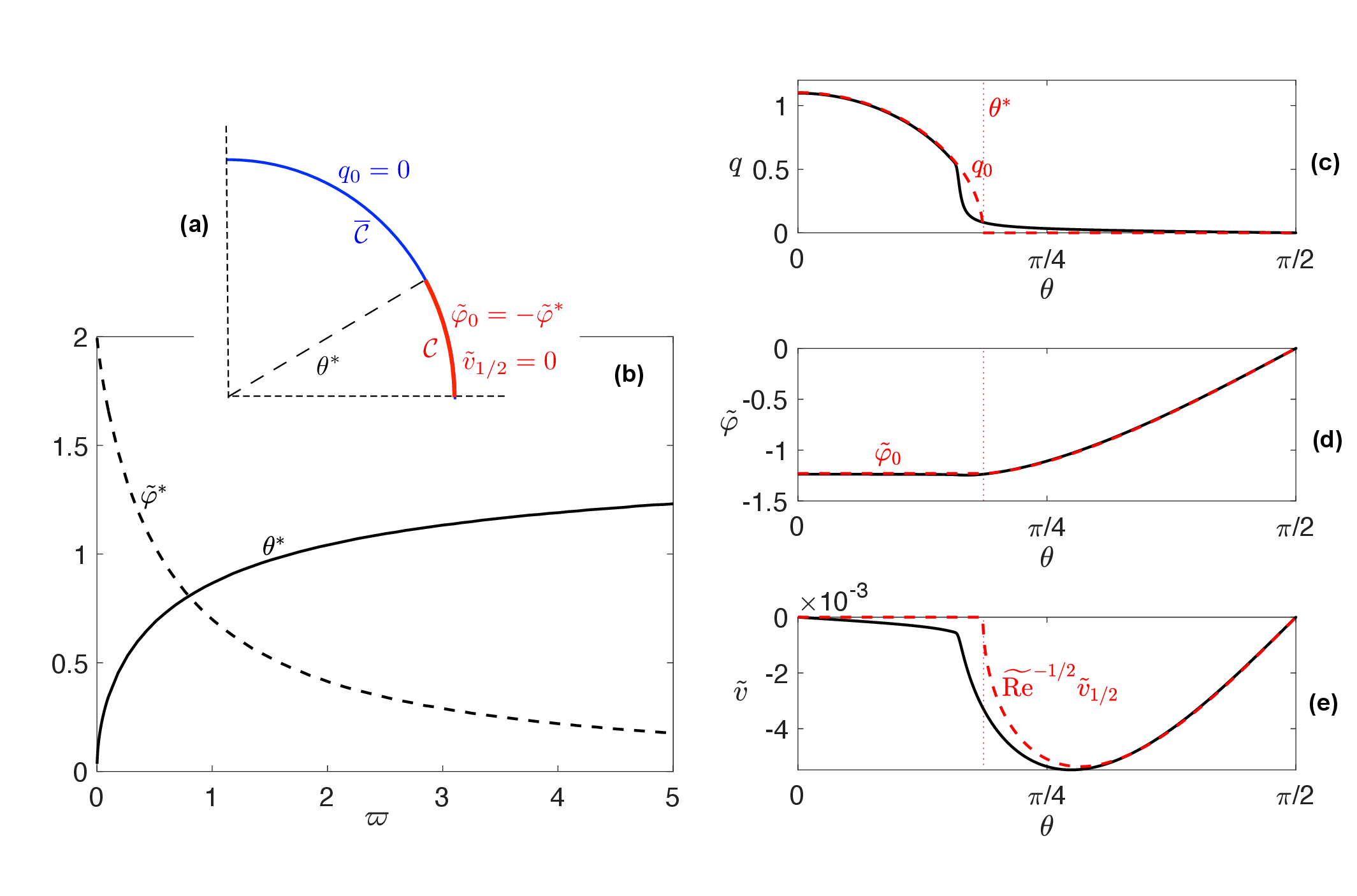}
\caption{Asymptotic theory for $\widetilde{\mathrm{Re}}\gg1$ in the parallel-polarization case $\varpi>0$. (a) Asymptotic decomposition of the first quadrant of the drop interface into a cap part $\mathcal{C}$ and its complement $\overline{\mathcal{C}}$; (b) Cap half-angle $\theta^*$ and voltage $\varphi^*$ as a function of $\varpi$; (c) Charge density $q$; (d) surface potential $\tilde{\varphi}$; and (e) surface velocity $\tilde{v}$ for $\varpi=1/3$ and $\widetilde{\mathrm{Re}}=1000$ --- comparison between full numerics (solid curves) and leading-order asymptotic approximations (dashed curves).}
\label{fig:cap_asymptotics}
\end{center}
\end{figure}

The interior potential $\tilde{\varphi}^{-}_0$ satisfies Laplace's equation in $r<1$, together with regularity at $r=0$ and mixed boundary conditions at $r=1$. The scaling arguments in subsection \ref{ssec:cap_structure} imply the uniform-potential condition 
\begin{equation}\label{cap phi0 bc}
\tilde{\varphi}_0=-\tilde{\varphi}^* 
\quad \text{on} \quad \mathcal{C},
\end{equation}
where we shall refer to the constant $\tilde{\varphi}^*$ as the cap voltage, 
and the zero-charge condition
\begin{equation}\label{cap q0 bc}
q_0  = 0  \quad \text{on} \quad \overline{\mathcal{C}}.
\end{equation}
The mixed boundary conditions \eqref{cap phi0 bc} and \eqref{cap q0 bc} are depicted in Fig.~\ref{fig:cap_asymptotics}a. 
Upon substitution into \eqref{q0 and phi0}, condition \eqref{cap q0 bc} is recast as the inhomogeneous Neumann condition
\begin{equation}\label{cap q0 bc explicit}
\pd{\tilde{\varphi}_0^-}{r}=-2\cos\theta  \quad \text{on} \quad \overline{\mathcal{C}}.
\end{equation}
Furthermore, $q_0$ and $\tilde{\varphi}_0^-$ satisfy symmetry conditions at $\theta=0$ and $\pi/2$ [cf.~\eqref{sym conditions poles} and \eqref{sym conditions equator}]. 

The above ``electric cap problem'' leaves both $\varphi^*$ and $\theta^*$ undetermined. In Appendix \ref{app:regularity}, we carry out a local analysis near the cap edge and invoke matching considerations to show that $q_0$ and $\partial\tilde{\varphi}_0/\partial\theta$ continuously vanish at the cap edge:
\refstepcounter{equation}
$$
\label{cap regularity}
\lim_{\theta\to\theta^*}q_0=0, \quad \lim_{\theta\to\theta^*}\pd{\tilde{\varphi}_0}{\theta}=0,
\eqno{(\theequation \mathrm{a},\mathrm{b})}
$$
with $q_0$ and $\partial\tilde{\varphi}_0/\partial{\theta}$ locally scaling as $\varrho^{1/2}$ on ${\mathcal{C}}$ and $\overline{\mathcal{C}}$, respectively, wherein $\varrho$ is the vanishing distance from the cap edge. 
For a given cap half-angle $\theta^*$, either of the (equivalent) regularity conditions \eqref{cap regularity}  ``closes'' the electric cap problem governing $\tilde{\varphi}_0^-$ (including the cap voltage $\tilde{\varphi}^*$). The half-angle $\theta^*$ itself is determined from the leading-order balance of the quadrant-charge condition \eqref{q quad regular}, 
\begin{equation}\label{net charge 0}
\int_{0}^{\theta^*}q_0\,\mathrm{d}\theta = \frac{2\varpi}{1+\varpi}.
\end{equation}

The parameters $\tilde{\varphi}^*$ and $\theta^*$ are functions of $\varpi$ alone. For given $\theta^*$, the electric cap problem is linear. To solve it we represent $\tilde{\varphi}_0^-$ by an appropriate Fourier-series solution of Laplace's equation [cf.~\eqref{pot fourier}] and use a collocation method to satisfy the mixed boundary conditions. In practice, on $\mathcal{C}$ we apply the uniform-potential condition $\partial{\tilde{\varphi}}_0/\partial \theta=0$, which is equivalent to \eqref{cap phi0 bc}. We find that our  numerical scheme then naturally converges to the unique solution satisfying  \eqref{cap regularity}, furnishing the correct value for $\tilde{\varphi}^*$ as a byproduct. Using this scheme as a mapping from $\theta^*$ to $q_0$, we obtain $\theta^*$ (and hence $q_0$, $\tilde{\varphi}_0^-$ and $\tilde{\varphi}^*$) as a function of $\varpi$ by numerically solving the nonlinear equation \eqref{net charge 0} using the \texttt{fsolve} solver in Matlab   \cite{MATLAB:R2022b}.

In Fig.~\ref{fig:cap_asymptotics}b, we show the variation of $\theta^*$ and $\tilde{\varphi}^*$ with $\varpi$. The cap half-angle $\theta^*$ increases monotonically with $\varpi$, vanishing as $\varpi\searrow0$ and approaching $\pi/2$ as $\varpi\to\infty$. In Figs.~\ref{fig:cap_asymptotics}c,d, we choose $\varpi=1/3$ and plot the surface potential $\tilde{\varphi}_0$ and charge $q_0$ against full numerics. In Fig.~\ref{fig:cap_stream}, we show electric field lines based on $\tilde{\varphi}^{\pm}_0$ for two values of $\varpi$; for the sake of comparison, we also depict the corresponding electric field lines for $\widetilde{\mathrm{Re}}=0$ [cf.~(\ref{zero Re solution tilde}c)]. 

\subsection{Hydrodynamic cap problem}
\label{ssec:cap_hydrodynamic}
We next formulate a problem governing the leading-order interior velocity field $\tilde{\bu}^-_{1/2}=\tilde{u}_{1/2}^-\be_r+\tilde{v}^-_{1/2}\be_{\theta}$. [The corresponding exterior velocity field $\tilde{\bu}^+_{1/2}$ can be obtained from $\tilde{\bu}^-_{1/2}$ using the reflection relation \eqref{psi reflection}.] It satisfies the Stokes equations for $r<1$, regularity at $r=0$, the impermeability boundary condition, 
\begin{equation}
\tilde{u}^-_{1/2}=0 \quad \text{at} \quad r=1,
\end{equation}
and mixed boundary conditions at $r=1$. The latter consist of the no-slip condition
\begin{equation}\label{no slip v half}
\tilde{v}_{1/2}=0 \quad \text{on} \quad \mathcal{C},
\end{equation}
which follows from the scaling analysis in Sec.~\ref{ssec:cap_structure}, 
and the tangential-stress balance
\begin{equation}\label{stress v half}
\pd{\tilde{v}_{1/2}^-}{r}-\tilde{v}_{1/2}=-q_{1/2}\pd{\tilde{\varphi}_0}{\theta} \quad \text{on} \quad \overline{\mathcal{C}},
\end{equation}
which follows from a leading-order balance of (\ref{factored equations}b).  Furthermore, the flow components $\tilde{u}^-_{1/2}$ and $\tilde{v}^-_{1/2}$ satisfy symmetry conditions at $\theta=0$ and $\pi/2$ [cf.~\eqref{sym conditions poles} and \eqref{sym conditions equator}]. 

\begin{figure}[t!]
\begin{center}
\includegraphics[scale=0.5,trim={15cm 3.5cm 18cm 2cm}]{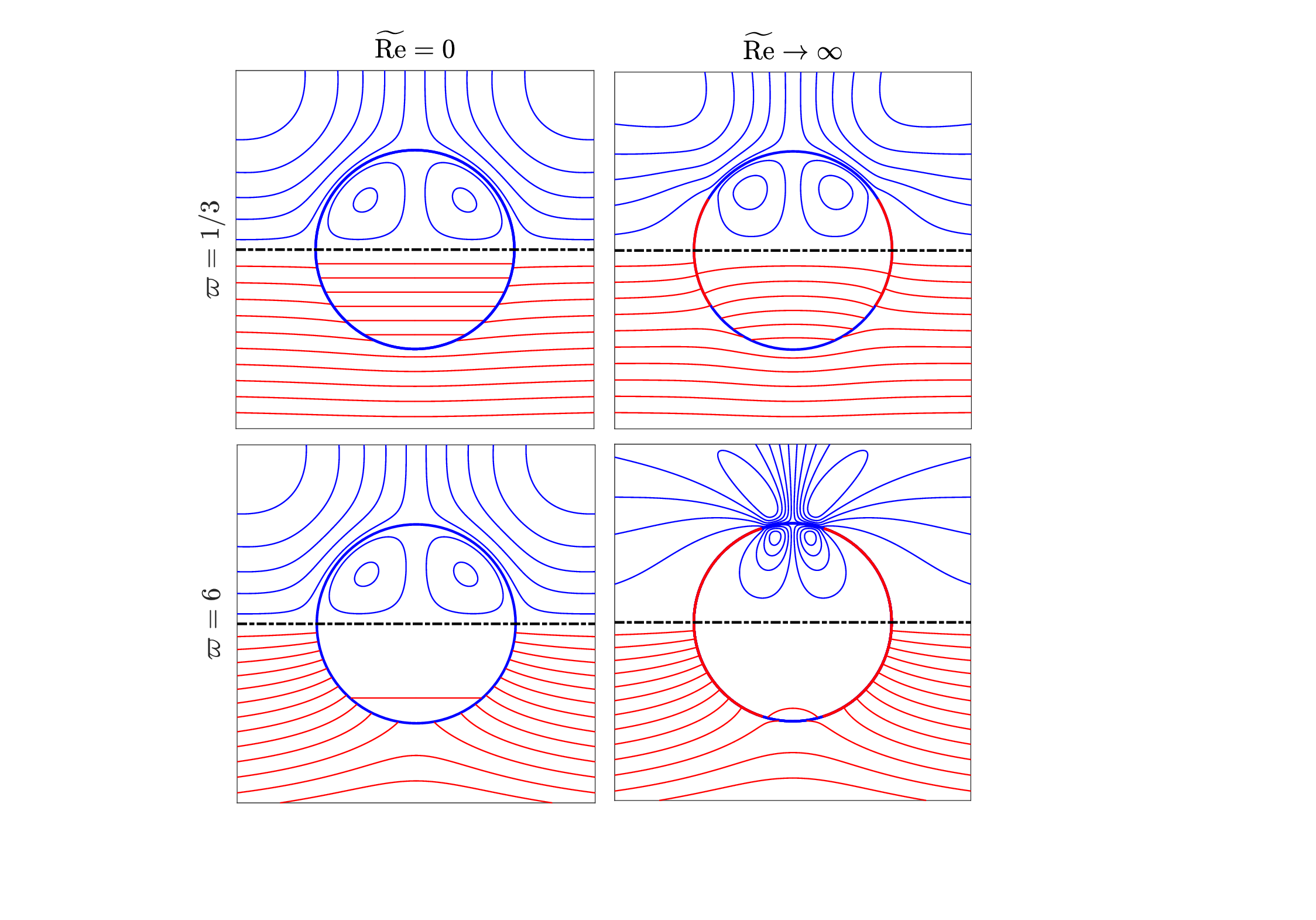}
\caption{Streamlines (upper half-planes) and electric field lines (lower half-planes) in the parallel-polarization cases $\varpi=1/3$ and $\varpi=6$, for $\widetilde{\mathrm{Re}}=0$ [cf.~\eqref{zero Re solution tilde}] and based on the large-$\widetilde{\mathrm{Re}}$ asymptotic scheme of Sec.~\ref{sec:caps}. The exterior field lines are calculated for the permittivity ratio $\mathcal{S}=1/3$.} 
\label{fig:cap_stream}
\end{center}
\end{figure}
In \eqref{stress v half}, $\partial{\tilde{\varphi}_0}/\partial\theta$ is known from the solution to the electric cap problem, whereas $q_{1/2}$ is unknown. Nonetheless, $q_{1/2}$ can be related to $v_{1/2}$ on $\overline{\mathcal{C}}$. The order-$\widetilde{\mathrm{Re}}^{-1}$ balance of the charging condition \eqref{charging in q} gives 
\begin{equation}\label{charging 1 on Cbar}
\pd{}{\theta}\left(q_{1/2}\tilde{v}_{1/2}\right)=2\varpi\cos\theta \quad \text{on} \quad \overline{\mathcal{C}},
\end{equation}
where we have used the fact that $q_0=0$ on $\overline{\mathcal{C}}$.
Thus, integrating \eqref{charging 1 on Cbar} and using the symmetry conditions (\ref{sym conditions equator}d) at $\theta=\pi/2$, we find 
\begin{equation}\label{vq E result}
\tilde{v}_{1/2}q_{1/2}=2\varpi(\sin\theta-1) \quad \text{on} \quad \overline{\mathcal{C}}.
\end{equation}
Combining \eqref{stress v half} and \eqref{vq E result}, we find the nonlinear boundary condition 
\begin{equation}\label{cap nonlinear stress}
\tilde{v}_{1/2}\left(\pd{\tilde{v}_{1/2}^-}{r}-\tilde{v}_{1/2}\right)=2\varpi(1-\sin\theta)\pd{\tilde{\varphi}_0}{\theta} \quad \text{on} \quad \overline{\mathcal{C}}.
\end{equation}

The above ``hydrodynamic cap problem'' is nonlinear and of the mixed boundary-value type. Like the electric cap problem, it depends solely on the parameter $\varpi$. Just as the asymptotic flow problem in the parallel-polarization scenario (Sec.~\ref{ssec:universalflow}), the present flow problem is also invariant under reversal of the velocity field. Similar to that scenario, the correct sign of the velocity field can be determined based on polarity; thus, \eqref{q sign positive}  together with \eqref{vq E result} implies  
\begin{equation}
\label{cap flow sign}
\tilde{v}_{1/2} < 0 \quad \text{on} \quad \overline{\mathcal{C}},
\end{equation} 
and so $\tilde{v}_{1/2}\le0$ for $0<\theta<\pi/2$, 
consistently with the schematic picture in Fig.~\ref{fig:scenarios}. Owing to the implicit dependence of $\theta^*$ upon $\varpi$, the hydrodynamic cap problem cannot be reduced to a ``universal'' one, as done for the asymptotic flow problem in the antiparallel-polarization scenario. Note that the dissipation argument, used originally in the antiparallel-polarization case (see subsection \ref{ssec:blowup_asym_scalings}), is consistent with \eqref{cap nonlinear stress} and $\varpi>0$. (The comparable contribution to the power integral on $\mathcal{C}$ vanishes owing to \eqref{no slip v half}.)

The solution of the hydrodynamic cap problem is implemented by representing the stream function corresponding to $\tilde{\bu}_{1/2}^-$ by an appropriate Fourier-series solution of the biharmonic equation [cf.~\eqref{psi fourier}] and using a collocation method to satisfy the mixed boundary conditions in conjunction with the \texttt{fsolve} solver in Matlab \cite{MATLAB:R2022b}. In Fig.~\ref{fig:cap_asymptotics}e, the profile of the azimuthal surface-velocity component $\tilde{v}_{1/2}$ is compared to full numerics for $\varpi=1/3$ and $\widetilde{\mathrm{Re}}=1000$. In Fig.~\ref{fig:cap_stream}, we show streamlines based on the flow $\tilde{\bu}_{1/2}^-$ for two values of $\varpi$; for the sake of comparison, we also show the corresponding field lines for $\widetilde{\mathrm{Re}}=0$ [cf.~(\ref{zero Re solution tilde}a)].

Despite the nonlinear boundary condition \eqref{cap nonlinear stress} being inhomogeneous, the local behaviour of the flow near the cap edge is readily seen to correspond to that conventionally expected near a sharp transition from no-slip to no-shear conditions \cite{Leal:book} such that the surface velocity $\tilde{v}_{1/2}$ on $\overline{\mathcal{C}}$ scales as $\varrho^{1/2}$ and the tangential stress $\partial\tilde{v}_{1/2}^-/\partial r-\tilde{v}_{1/2}$ scales as $\varrho^{-1/2}$ on $\mathcal{C}$, $\varrho$ being the vanishing distance from the cap edge. Indeed, this homogeneous local behavior dominates that implied by the inhomogeneous term on the right-hand side of \eqref{cap nonlinear stress}, corresponding to $\tilde{v}_{1/2}$ on $\overline{\mathcal{C}}$ scaling as $\varrho^{3/4}$. 

Having determined the local behavior of the outer flow field, in addition to the local behaviors of the corresponding charge density and electric potential [see \eqref{cap regularity} and Appendix \ref{app:regularity}], we can now comment on the need for the hypothesized inner region near the cap edge. On one hand, since the outer tangential viscous stress at order $\widetilde{\mathrm{Re}}^{-1/2}$ diverges like $\varrho^{-1/2}$ as $\varrho\searrow0$ on ${\mathcal{C}}$, we expect an $\text{ord}(\widetilde{\mathrm{Re}}^{-1/2}\varrho^{-1/2})$ tangential viscous stress near the cap edge. On the other hand, with the outer charge density and tangential electric field, both of order unity, vanishing like $\varrho^{1/2}$ as $\varrho\searrow0$ on $\mathcal{C}$ and $\overline{\mathcal{C}}$, respectively, we expect an order-$\varrho$ tangential Maxwell stress near the cap edge. Upon balancing these stress estimates, we find that the dominance of the Maxwell stress on $\mathcal{C}$ (associated with the  caps being perfectly conducting) breaks down for $\varrho=\mathcal{O}(\widetilde{\mathrm{Re}}^{-1/3})$, indicating an inner region of that width. The local behaviors of $q_0$ and $\tilde{v}_{1/2}$ as $\varrho\searrow0$ then imply that $q=\mathcal{O}(\widetilde{\mathrm{Re}}^{-1/6})$ and $\tilde{v}=\mathcal{O}(\widetilde{\mathrm{Re}}^{-2/3})$ in that region \footnote{Substituting these scalings into the charging equation \eqref{charging in q} shows that surface convection remains dominant on the inner scale --- we therefore anticipate an intricate asymptotic structure involving ``nested'' inner regions, with the smoothness of the solutions only becoming discernible on some scale  smaller than the inner one.}.

\section{Discussion}
\label{sec:discussion}
\subsection{Taylor-symmetric steady state}
We have considered the two-dimensional electrohydrodynamics problem of a circular drop in an external electric field, focusing on the ``Taylor-symmetric'' steady state that continues the unique weak-field solution to arbitrary values of the electric Reynolds number $\mathrm{Re}_E$. In formulating the problem governing that steady state, we have identified a ``factorization'' of the governing fields that effectively reduces the number of dimensionless parameters from four --- $\mathrm{Re}_E$ and the fluid ratios $\mathcal{R}$, $\mathcal{S}$ and $\mathcal{M}$ --- to just two: the modified electric Reynolds number $\widetilde{\mathrm{Re}}$ and charging parameter $\varpi$. This factorization, which relies on the Taylor symmetry, is based on inversion of both the stream function and the (``corrected'') electric potential about the circular drop interface. As such, it has the added benefit of reducing the problem domain to the union of the drop interior and its interface. 

We have shown that the Taylor-symmetric steady state possesses a general ``polarity'' property, which is familiar from weak-field theory but turns out to hold for all $\widetilde{\mathrm{Re}}$: If charge relaxation in the drop phase is slower than in the suspending phase ($-1<\varpi<0$), the interface polarizes antiparallel to the external field such that the charge density is positive over its posterior half and negative over its anterior half. Conversely, if the suspending phase relaxes slower ($\varpi>0$), the interface polarizes parallel to the external field such that the charge density is negative over its posterior half and positive over its anterior half. With the understanding that the effects of surface-charge convection are very different in these two scenarios, we have separately analyzed the scenarios of antiparallel and parallel polarization.

\subsection{Equatorial blowup}
The most notable consequence of surface-charge convection in the antiparallel-polarization case ($-1<\varpi<0$) is the formation of a singularity at the drop equator, namely at the two interfacial points across which the charge density flips sign. As discussed in the Introduction, the steady state of the analogous three-dimensional scenario could not be attained for a sufficiently strong external field in previous numerical studies. In contrast, we have herein used a combination of numerical and asymptotic analysis to effectively continue the steady state deep into the singular regime --- essentially to arbitrary $\widetilde{\mathrm{Re}}$. 

Let us summarize the evolution of the equatorial singularity with increasing $\widetilde{\mathrm{Re}}$ as it is understood from our investigations. For sufficiently small $\widetilde{\mathrm{Re}}$, the solution is smooth, with the slope at the equator of the normalized charge density $q$ steepening with increasing $\widetilde{\mathrm{Re}}$. As $\widetilde{\mathrm{Re}}$ approaches a first critical value, which depends upon $\varpi$, that slope diverges. For larger $\widetilde{\mathrm{Re}}$, the continuation of the solution is no longer smooth. At first, $q$ remains continuous, behaving sublinearly like $|\Delta \theta|^{\alpha}\mathrm{sgn}\Delta\theta$ as the angle from the equator $\Delta\theta$ vanishes, with the exponent $\alpha$ decreasing monotonically from unity, apparently towards zero, with increasing $\widetilde{\mathrm{Re}}$. Using exact local arguments, we have explicitly related both the slope steepening in the smooth regime and the subsequent exponent decrease in the transition regime to the concurrent steepening of the equatorial slope of the normalized surface velocity $\tilde{v}$. In both regimes, $\tilde{v}$ behaves like $\Delta\theta$ near the equator, with the local flow simply extrapolating that driven globally by electric stresses over the entire drop interface.

Above a second critical value of $\widetilde{\mathrm{Re}}$ (also dependent upon $\varpi$), the solution exhibits a blowup singularity where $q$ asymmetrically diverges like $|\Delta \theta|^{-1/3}\mathrm{sgn} \Delta\theta$ as $\Delta\theta\to0$ --- corresponding to a \emph{fixed} negative exponent $\alpha=-1/3$ (independent of $\widetilde{\mathrm{Re}}$ and $\varpi$). Furthermore, $\tilde{v}$ vanishes like $|\Delta \theta|^{1/3}\mathrm{sgn} \Delta\theta$, whereby the normalized surface current $q\tilde{v}$ possesses a finite non-zero limit as $\Delta\theta\to0$, rather than continuously vanishing as in the preceding smooth and transition regimes. In contrast to those regimes, the leading-order flow, charge density and electric field near the equator are locally induced. We have employed local analysis to derive the corresponding singularity structure in detail, finding it to be characterized by a single blowup prefactor which cannot be determined solely from local considerations. Physically, the blowup singularity  describes opposite surface charges being convected by a locally induced flow towards the equator, where they ``annihilate.'' While charging of the interface via bulk Ohmic currents is locally negligible in this singularity structure, the blowup prefactor $A$ is globally determined by a balance between Ohmic charging over the drop and equatorial annihilation. 

Numerically, we have used a semi-analytical scheme based on Fourier-series expansions. To extend this scheme into the blowup regime, we have employed singularity reduction by encoding the leading-order local singularity structure directly in Fourier space, with the global charging--annihilation balance used as an auxiliary condition to determine the blowup prefactor as part of the solution. We have thereby calculated steady solutions up to large values of $\widetilde{\mathrm{Re}}$, demonstrating that as a function of $\widetilde{\mathrm{Re}}$ the blowup prefactor $A$ bifurcates from zero and subsequently varies non-monotonically; and that at the blowup threshold the scaled quadrant charge $\mathcal{Q}$ bifurcates from its fixed pre-blowup value and subsequently monotonically decreases in absolute magnitude. We note that the present numerical scheme struggles very close to the blowup threshold; in particular, we cannot accurately compute small values of the exponent $\alpha$ in the transition interval as the blowup threshold is approached from below, nor the precise location of that threshold.  

We have supplemented our numerical solutions by deriving asymptotic approximations in the limit $\widetilde{\mathrm{Re}}\gg1$, corresponding to a regime far beyond the blowup threshold. We have found the charge density $q$ (and quadrant charge $\mathcal{Q}$) and velocity fields $\tilde{\bu}^{\pm}$ to scale as $\widetilde{\mathrm{Re}}^{-1/2}$. Given the smallness of $q$, the leading-order potential distribution corresponds to that familiar from elementary electrostatic theory in the case of perfect dielectric fluids. The leading-order balance of the interfacial charging equation can then be integrated analytically, furnishing an explicit leading-order approximation for the surface current, namely the product of the surface-charge density and surface velocity. Combining that result with the flow problem then leads, upon normalization, to a parameter-free nonlinear boundary-value problem, namely a Stokes problem wherein the flow is forced by a prescribed power density over the interface. We have solved this problem numerically, obtaining universal approximations for the velocity field (and charge density) on which numerical solutions of the full problem collapse at large $\widetilde{\mathrm{Re}}$. 

It is worth reiterating two subtleties involved in the large-$\widetilde{\mathrm{Re}}$ analysis. The first has to do with the local behaviors of the universal solutions near the equator. While the approximation for the charge density  diverges as the equator is approached, it does so faster than expected from the blowup structure obtained by local analysis of the exact problem at fixed $\widetilde{\mathrm{Re}}$ --- specifically, like $|\Delta\theta|^{-1/2}$ rather than $|\Delta\theta|^{-1/3}$. The resolution of that discrepancy  lies with the spatial non-uniformity of the asymptotic limit $\widetilde{\mathrm{Re}}\gg1$: the universal solutions correspond to an outer drop-scale approximation, which fails in an $\mathcal{O}(1/\widetilde{\mathrm{Re}})$ vicinity of the equator. Nonetheless, the strength of the ``true'' blowup singularity can  be extracted directly from the outer approximation, circumventing the need for a detailed inner analysis of that vicinity. Indeed, the global charging--annihilation balance gives the blowup prefactor $A$ as a function of the net quadrant charge $\mathcal{Q}$, which is dominated by the outer contribution. In this manner, we have obtained a two-term expansion for $A$, starting at order $\widetilde{\mathrm{Re}}^{-1/3}$, as well as a leading-order approximation for $\mathcal{Q}$, which scales as $\widetilde{\mathrm{Re}}^{-1/2}$, similarly to the charge density. Furthermore, we have shown that the asymptotic expansions for $A$ and $\mathcal{Q}$ can be recast as universal approximations on which numerical data collapse at large $\widetilde{\mathrm{Re}}$.

The second subtlety is that the nonlinear boundary-value problem governing the universal flow approximation is invariant under reversal of the velocity field. From the corresponding surface-current approximation, such a reversal is coupled to a sign reversal of the charge density. Fortunately, the latter linkage allowed us to determine the correct sign of the velocity field by alluding to the global charge-polarity result. Without it, we would have had to resort to matching with the inner region near the equator, where the local blowup structure demands a surface flow converging towards the equator. 

It is instructive to compare the present large-$\widetilde{\mathrm{Re}}$ analysis with that of \citet{Yariv:16}, who considered the same two-dimensional setup of a circular drop in an electric field but focused on Quincke-rotation steady states wherein the Taylor-symmetry is broken. Remarkably, our scalings for the charge density and flow field are the same as theirs, as are the leading-order approximations for the electric field and surface current. In turn, the leading-order flow problems are essentially the same; the present one, in fact, being a restricted variant of theirs owing to the reflection of the stream function about the drop interface --- a simplification inapplicable to the Quincke-rotation scenario where a slowly decaying irrotational vortex is present in the exterior domain. Despite the similarities in the asymptotic structure, the resulting approximations for the charge density and flow are, of course, very different in the Taylor-symmetric and symmetry-broken scenarios. In particular, the Taylor-symmetric approximations are singular at the equator and asymptotically nonuniform there, whereas the symmetry-broken approximations are smooth and asymptotically uniform. Furthermore, while the invariance of the leading-order flow problem under flow reversal is removable in the Taylor-symmetric scenario by the polarity result, in the Quincke-rotation scenario it represents the arbitrariness of the rotation direction \cite{Yariv:16}. 

As pointed out by Yariv \& Frankel \cite{Yariv:16}, the $\widetilde{\mathrm{Re}}^{-1/2}$ scaling of the velocity field and charge density means that, in dimensional terms, the flow scales with the first power of the external-field magnitude, while the charge density is, remarkably, independent of that magnitude. This contrasts the respective quadratic and linear scalings familiar from weak fields.

\subsection{Polar caps}

In the parallel-polarization case ($\varpi>0$), the field acting on its induced surface charge drives a flow from the equator to the poles, thus convecting positive and negative charges towards the front and back poles, respectively. In that scenario, our numerical solutions suggest that no singularity arises regardless of the value of $\widetilde{\mathrm{Re}}$ --- in marked contrast to the antiparallel-polarization case. Rather, the key consequence of surface-charge convection in the parallel-polarization case is that at large $\widetilde{\mathrm{Re}}$ surface charge is confined to two ``caps'' centered about the drop poles, which are hydrodynamically stagnant as well as electrically perfectly conducting. We have illuminated this novel phenomenon through numerical simulations and asymptotic analysis in the limit $\widetilde{\mathrm{Re}}\to\infty$. 

The asymptotic scheme, which is fundamentally different from that in the antiparallel-polarization case, results in a pair of electric and hydrodynamic mixed boundary-value problems, both of which involve $\varpi$ as the sole parameter. The ``electric cap problem,'' which is decoupled from the flow, linearly determines the order-unity potential field (and cap voltage, in particular) for a given cap angle. That angle, in turn, is nonlinearly determined by a global charging balance which effectively fixes the net cap charge. From the numerical solution of the electric problem we find that the cap angle grows and the cap voltage decreases monotonically with increasing $\varpi$ (increasing conductivity or decreasing permittivity of the drop phase relative to the suspending phase). The ``hydrodynamic cap problem'' then determines the flow, of order $\widetilde{\mathrm{Re}}^{-1/2}$, which is nonlinearly driven by electrical shear stresses at the cap complement. This contrasts classical surfactant-transport problems at large surface P\'eclet numbers featuring the formation of stagnant caps, where the flow is externally driven and satisfies a no-shear condition there \cite{Leal:book}. 

As in the parallel-polarization case, the $\widetilde{\mathrm{Re}}^{-1/2}$ scaling of the velocity field means that, in dimensional terms, the flow scales anomalously, with the first power of the external-field magnitude, contrasting the quadratic scaling with that magnitude familiar from weak-field theory. Unlike in the parallel-polarization case, however, the charge density is of order unity (in the cap regions), implying that dimensionally it scales linearly with the  external-field magnitude as it does in the weak-field regime. 

Despite the numerical solutions appearing to be smooth in the parallel-polarization case, the cap regions in the large-$\widetilde{\mathrm{Re}}$ asymptotic scheme terminate sharply; accordingly, the solutions to the electric and hydrodynamic cap problems are only piecewise smooth over the drop interface. We have conjectured that these solutions correspond to leading-order outer approximations describing the drop scale, and that smoothness of the physical fields could be demonstrated via asymptotic matching with inner regions near the cap edges. While such an analysis is outside the scope of this work, we have alluded to the existence of such inner regions in deriving the cap-edge regularity condition that closes the electric cap problem. In the absence of surface-charge diffusion, the small-scale smoothing of the solutions near the cap edges must necessarily arise from nonlocal coupling of the interfacial charge transport to the bulk domains. In the context of classical studies of surfactant-cap formation, such a nonlocal smoothing mechanism is analogous to cap-edge smoothing via kinetic transport at infinite surface P\'eclet number \cite{Wang:99}, rather than the much more trivial case where the cap-edge smoothing is via surface diffusion at large surface P\'eclet numbers \cite{Harper:92}.  

\section{Concluding remarks}
\label{sec:conclusions}
To conclude, we have analyzed the symmetric base state of a drop in an electric field assuming a two-dimensional leaky-dielectric model  wherein the drop is a non-deformable disk. This simplified framework  facilitated straightforward numerics and enabled a surprising factorization of the problem formulation, effectively reducing the number of dimensionless parameters by two. An obvious follow-up of the present work would be to analyze the analogous three-dimensional --- in fact, axisymmetric --- base state, namely the continuation of Taylor's weak-field solution for a spherical drop to arbitrary electric Reynolds numbers. 

In hindsight, we expect the axisymmetric problem to be similar in several aspects. For a start, we anticipate analogous fore-aft symmetries and linkage between polarity and the ratio of relaxation time scales. We accordingly hypothesize that surface-charge convection  results in essentially the same fundamental phenomena, namely equatorial singularity formation in the antiparallel-polarization case, starting at moderate electric Reynolds numbers, and the formation of polar caps in the parallel-polarization case, at large electric Reynolds numbers. In relation to the former phenomenon, the fact that the axisymmetric problem effectively reduces to a planar one near the equator suggests that the local blowup structure is the same as herein. Furthermore, we anticipate that the asymptotic structure at large electric Reynolds numbers is similar to that found herein in both the parallel- and antiparallel-polarization cases. On a technical note, Kelvin inversion for potentials and axisymmetric Stokes flow \cite{Dassios:09} could be used to reduce the problem domain to the drop interior and its boundary; in the axisymmetric problem, however, such reflection about the drop interface can be shown to eliminate only one dimensionless parameter (the viscosity ratio). 

In retrospect, it is clear that the blowup singularity structure is universal to the leaky-dielectric equations. Indeed, we have herein developed this structure (up to a single blowup prefactor) by means of a local analysis near the drop equator, assuming only the symmetries inherited from the global formulation and that the charge density diverges approaching the equator. Besides the axisymmetric drop problem discussed above, this singularity is also expected to form in the menisci \cite{Malkus:61} and thin-film \cite{Jolly:70} setups studied in the 1960s in the context of the surface-electroconvection instability; the periodic problem considered by  \citet{Firouznia:21}; and drop electrocoalscence \cite{Chen:23}. In fact, said universality holds even if accounting for deformation: the local blowup structure is associated with only weakly singular electric and viscous normal stresses, both scaling with the $-2/3$ power of distance from the singularity --- too weak to induce a locally appreciable deformation from a flat interface. Thus, the local blowup structure could be subtracted in steady-state numerical simulations of deformable interfaces, for example towards studying the drop problem beyond small electric capillary numbers. 

Given its universality, it is instructive to describe the steady blowup structure dimensionally and in a manner detached from the specifics of the drop problem. To this end, consider the local behavior approaching a surface curve on an interface between two leaky-dielectric liquids. Let $s_*$ be an arc-length surface coordinate that changes sign upon crossing that surface curve. Assuming steady blowup occurs at the surface curve, the surface velocity is locally normal to it, directed towards it from both sides. The charge density $q_*$ changes sign upon crossing the surface curve; without loss of generality we assume $q_*<0$ for $s_*>0$. Furthermore, the surface current in the direction of increasing $s_*$ possesses a finite double-sided limit, which we denote by $J_*$. With these conventions, 
\begin{equation}\label{dimensional singularity}
q_*\sim-\left(\frac{2J_*}{3}\right)^{1/3}(\epsilon^+_*+\epsilon^-_*)^{1/3}(\mu^+_*+\mu^-_*)^{1/3}|s_*|^{-1/3}\sgn s_*  \quad \text{as} \quad s_*\to0,
\end{equation} 
while the surface velocity in the direction of increasing $s_*$ can be obtained as $u_*\sim J_*/q_*$. We note that the liquid permittivities $\epsilon_*^{\pm}$ and viscosities $\mu_*^{\pm}$ appear in a symmetric fashion, while the absence of the liquid conductivities reflects the fact that surface convection locally dominates charge relaxation.  

The present study has been entirely focused on steady solutions. As discussed in the Introduction, however, singularity formation in the problem of a drop in an electric field was initially observed in \emph{unsteady} numerical simulations in which the interface is initially uncharged \cite{Lanauze:15,Das:17,Firouznia:23}. Given the singular nature of the symmetric steady state for sufficiently large electric Reynolds number, we hypothesize that the apparent shock-like charge-density profiles observed in those simulations (just before numerical failure) correspond to snapshots of the solution's evolution towards a finite-time \emph{blowup}  singularity. We are currently investigating the possibility of self-similar unsteady blowup in the leaky-dielectric model and connections to the steady singular solutions studied herein. 

Lastly, a key motivation for the present study has been to lay the foundations for studying the stability of the base state of a drop in an electric field, which we believe involves the interplay between singularity formation and the Quincke and surface-electroconvection instability mechanisms. In light of the singular nature of the base state, a straightforward linearization may prove unsuccessful. One approach would be to regularize the underlying model. A physically accurate regularization could only be obtained by a systematic coarse graining of electrokinetic models \cite{Schnitzer:15,Mori:18,Ma:22}. A simpler approach would be to incorporate weak surface-charge diffusion into the leaky-dielectric model, as done in some numerical works \cite{Wagoner:20,Wagoner:21}. An alternative approach would be to consider the stability of an unsteady base-state solution corresponding to an initially uncharged interface, which as mentioned above is expected to exhibit finite-time blowup. 

\textbf{Acknowledgements.} O.~Schnitzer and G.~G.~Peng acknowledge the support of the Leverhulme Trust through Research Project Grant RPG-2021-161. The authors also acknowledge discussions with Sara Drummond-Curtis and Demetrios Papageorgiou on the possibility of using reflection in order to factorize the analogous axisymmetric problem as discussed in Sec.~\ref{sec:conclusions}.

\appendix
\section{Polarity}
\label{app:polarity}
We argue that for a Taylor-symmetric steady state the sign of the charge density $q$ in the first and fourth quadrant, $-\pi/2<\theta<\pi/2$, must be equal to the sign of the charging parameter $\varpi$, thereby giving the polarization results \eqref{q sign positive} and \eqref{q sign negative}. (Since $q$ is antisymmetric about the equatorial line, its sign must then be opposite in the second and third quadrants.) We shall initially assume that $q$ and the surface velocity $\tilde{v}$ are smooth functions of $\theta$, and subsequently consider the possibility of singular behavior at the equator. 

We begin with an intuitive explanation. The charge $q$ satisfies the modified charging condition \eqref{charging in q}, 
\begin{equation}
\widetilde{\mathrm{Re}}\pd{}{\theta}\left(q\tilde{v}\right) = 2\varpi\cos\theta - (1+\varpi)q,
\label{polarity convection}
\end{equation}
which is a steady convection equation with a fixed source term $2\varpi\cos\theta$ that has the same sign as $\varpi$ for $-\pi/2< \theta < \pi/2$, and a decay term $-(1+\varpi)q$ proportional to $q$. Since $q$ and $\tilde{v}$ are antisymmetric about the equator, the surface current $q\tilde{v}$ vanishes at $\theta=\pm\pi/2$, so there is no convection of charge across the equator and we expect $q$ to have the same sign as the source. 

We now show this more carefully for the case $\varpi > 0$; the  same argument applies to the case $\varpi < 0$ with all signs reversed. We first show that $q \geq 0$ for $-\pi/2 < \theta < \pi/2$. Assume, on the contrary, that $q < 0$ at some $\theta$ in that range. It follows by continuity that $q < 0$ in some interval $\theta_1 < \theta < \theta_2$, and we choose this interval to be maximal, so that $q=0$ at $\theta = \theta_{1,2}$. (Since $q$ is antisymmetric about the equator, the interval is bounded: $-\pi/2 \leq \theta_1 < \theta_2 \leq \pi/2$.) Integrating \eqref{polarity convection} over this interval yields
\begin{equation}\label{polarity integrated}
\widetilde{\mathrm{Re}}\left[(q\tilde{v})|_{\theta=\theta_2} - (q\tilde{v})|_{\theta=\theta_1}\right] = 2\varpi (\sin\theta_2 - \sin\theta_1) - (1+\varpi) \int_{\theta_1}^{\theta_2} q \,\mathrm{d}\theta,
\end{equation}
where the right-hand side is strictly positive
 as it represents the contribution from the source and from the decay of a negative quantity, while the left-hand side which represents convection of charge across the boundaries $\theta_{1,2}$ of the interval vanishes, as the interval was chosen so that $q=0$ on the boundaries. This contradiction shows that $q \geq 0$ throughout $-\pi/2 < \theta < \pi/2$.
 
It remains to consider the possibility that $q$ vanishes at some $\theta_*$ in $-\pi/2 < \theta_* < \pi/2$. In that case, \eqref{polarity convection} yields $v(\theta_*) q'(\theta_*) = 2 \varpi \cos \theta_*$, from which it follows that $q'(\theta_*) \neq 0$ and hence that $q$ changes sign there, which contradicts our result that $q \geq 0$. Hence, $q$ must be strictly positive throughout $-\pi/2 < \theta < \pi/2$.

Consider now the possibility of the solution exhibiting some form of singularity at the equator $\theta=\pm\pi/2$ (as indeed turns out to be the case in the antiparallel-polarization case $-1<\varpi<0$ and sufficiently large $\widetilde{\mathrm{Re}}$). As discussed in Sec.~\ref{sec:formulation}, the physical  requirement for $q$ to be integrable and the global charging balance \eqref{q quad singular} implies the existence and finiteness of the surface-current limit  $\lim_{\theta\to\pi/2}q\tilde{v}=-\lim_{\theta\to-\pi/2}q\tilde{v}$ (note the antisymmetry of $q\tilde{v}$ about the pole line). As long as that limit vanishes, 
 our arguments above for the polarity result generalize trivially. 

If $\lim_{\theta\to\pm\pi/2}q\tilde{v}$ are non-zero, the polarity result can still be similarly argued in the case where the surface flow is locally towards the equator; as explained below \eqref{gunnar singular}, in that case the non-zero surface-current limit represents annihilation, rather than creation, of positive and negative charges. Indeed, revisiting the intuitive explanation, we still expect $q$ in $-\pi/2<\theta<\pi/2$ to have the same sign as the source as there is then no convection of opposite charge across the equator into the region $-\pi/2 < \theta < \pi/2$. 

The more careful argument only needs adjustment in the case where $\theta_1$ and/or $\theta_2$ cannot be chosen in the region $-\pi/2<\theta<\pi/2$. Consider, for example, the case where $\theta_1$ but not $\theta_2$ can be found in that region. Then from the assumption that $q<0$ for some $\theta$ in the region $-\pi/2 < \theta < \pi/2$ and continuity it follows that $q<0$ for some interval $-\pi/2<\theta_1<\theta<\theta_{\delta}<\pi/2$ such that (i) $q$ vanishes at $\theta_1$; and (ii) $\theta_\delta=\pi/2-\delta$, with $\delta>0$ sufficiently small so that $\left.\tilde{v}\right|_{\theta=\theta_\delta}>0$. Integrating \eqref{polarity convection} over $\theta_1<\theta<\theta_{\delta}$ gives a balance like \eqref{polarity integrated} but with $\theta_\delta$ replacing $\theta_2$. The right-hand side remains strictly positive while now the left-hand side is negative, rather than zero, since $(q\tilde{v})|_{\theta = \theta_\delta} < 0$. Thus, the contradiction follows as in the original argument.  

\section{Numerical scheme}
\label{app:numerics}
\subsection{Fourier-series scheme}
\label{app:numerics_fourier}
Consider the factorized formulation of subsection \ref{ssec:reduced}. Given the Taylor symmetry stated in subsection \ref{ssec:symmetry}, the surface-charge density $q$ possesses the Fourier-series representation 
\begin{equation}\label{q fourier}
q=\sum_{n=1}^{\infty}\alpha_n\cos((2n-1)\theta),
\end{equation}
where the coefficients $\alpha_n$ are real. In terms of these coefficients, the quadrant charge $\mathcal{Q}$ [cf.~\eqref{Q def}] can be expressed as 
\begin{equation}\label{quadrant charge fourier}
\mathcal{Q}=-\sum_{n=1}^{\infty}(-1)^n\frac{\alpha_n}{2n-1},
\end{equation}
while the problem for the interior potential $\tilde{\varphi}^-$, consisting of Laplace's equation for $r<1$, regularity at the origin and the displacement condition (\ref{factored equations}a), can be solved to give
\begin{equation}\label{pot fourier}
\tilde{\varphi}^-=-2r\cos\theta+\sum_{n=1}^{\infty}\frac{\alpha_n}{2n-1}r^{2n-1}\cos((2n-1)\theta),
\end{equation}
where the constant term vanishes in accordance with the Taylor symmetry. 
In particular, the azimuthal derivative at $r=1$ is
\begin{equation}\label{pot theta fourier}
\pd{\tilde{\varphi}}{\theta}=2\sin\theta-\sum_{n=1}^{\infty}\alpha_n\sin((2n-1)\theta). 
\end{equation}
The exterior potential $\tilde{\varphi}^+$ can be found from \eqref{pot fourier} using the reflection relation \eqref{pot reflection}.

Consider next the interior stream function $\tilde{\psi}^-$. A general solution of the biharmonic equation for $r<1$ that possesses the Taylor symmetry and satisfies regularity at the origin and impermeability at $r=1$ is 
\begin{equation}\label{psi fourier}
\tilde{\psi}^-=\sum_{n=1}^{\infty}\frac{\beta_n}{4(2n-1)}r^{2n}(r^{2}-1)\sin 2n\theta,
\end{equation}
where the coefficients $\beta_n$ are real. 
In particular, the surface azimuthal velocity and tangential stress are represented by the Fourier series 
\refstepcounter{equation}
\label{v and stress fourier}
$$
\tilde{v}=\sum_{n=1}^{\infty}\frac{\beta_n}{2(1-2n)}\sin 2n\theta, \quad \pd{\tilde{v}^-}{r}-\tilde{v}=\sum_{n=1}^{\infty}\frac{2n}{1-2n}\beta_n\sin 2n\theta. 
\eqno{(\theequation \mathrm{a},\mathrm{b})}
$$
The exterior stream function $\tilde{\psi}^+$ can be found from \eqref{psi fourier} using the reflection relation \eqref{psi reflection}.

It remains to satisfy the shear-stress condition (\ref{factored equations}b) and the charging condition \eqref{charging in q}. The right-hand side of condition (\ref{factored equations}b) has the Fourier-series representation 
\begin{equation}
-q\pd{\tilde\varphi}{\theta}=\sum_{n=1}^{\infty}\chi_n\sin2n\theta,
\end{equation}
where, using \eqref{q fourier} and \eqref{pot theta fourier}, we find 
\begin{equation}\label{chi coefficients}
\chi_n=-\alpha_n+\alpha_{n+1}+\frac{1}{2}\sum_{k=1}^{n}\alpha_{k}\alpha_{n+1-k}, \quad n=1,2,\ldots 
\end{equation}
Thus, condition (\ref{factored equations}b) gives 
\begin{equation}
\beta_n=\frac{1-2n}{2n}\chi_n, \quad n=1,2,\ldots,
\end{equation}
where we have used (\ref{v and stress fourier}b).

The charging condition \eqref{charging in q} can be integrated, using the symmetry condition (\ref{sym conditions poles}d) and \eqref{q fourier}, to give 
\begin{equation}\label{qv fourier}
q\tilde{v}=2\varpi\widetilde{\mathrm{Re}}^{-1}\sin\theta-\widetilde{\mathrm{Re}}^{-1}(1+\varpi)\sum_{n=1}^{\infty}\frac{\alpha_n}{2n-1}\sin((2n-1)\theta),
\end{equation}
The Fourier series \eqref{qv fourier} must be compatible with that obtained by multiplying \eqref{q fourier} and (\ref{v and stress fourier}a), 
\begin{equation}\label{qv fourier direct}
q\tilde{v}=\sum_{n=1}^{\infty}\gamma_n\sin ((2n-1)\theta),
\end{equation}
where
\begin{equation}\label{gamma coefficients}
\gamma_n=\sum_{k=1}^{n-1}\frac{\chi_k\alpha_{n-k}}{8k}-\sum_{k=1}^{\infty}\frac{\chi_{k}\alpha_{n+k}}{8k}+\sum_{k=1}^{\infty}\frac{\chi_{n+k-1}\alpha_k}{8(n+k-1)}, \quad n=1,2\ldots
\end{equation}
Thus, comparing \eqref{qv fourier} and \eqref{qv fourier direct} gives 
\begin{equation}\label{qv condition fourier}
-\widetilde{\mathrm{Re}}^{-1}(1+\varpi)\frac{\alpha_n}{2n-1}+2\varpi\widetilde{\mathrm{Re}}^{-1}\delta_{n,1}=\gamma_n, \quad n=1,2\ldots,
\end{equation}
where $\delta_{n,1}$ is the Kronecker Delta symbol. 
 
The Fourier coefficients $\alpha_n$ and $\chi_n$ are governed by the infinite system of equations \eqref{chi coefficients} together with the infinite system of equations obtained by substituting \eqref{gamma coefficients} into \eqref{qv condition fourier}. The preliminary results presented in Sec.~\ref{sec:preliminary} were obtained by numerically solving these equations by controlled truncation and the Matlab solver \texttt{fsolve} \cite{MATLAB:R2022b}; given the quadratic nonlinearities, analytical expressions for the Jacobian are readily obtained. In the case of smooth solutions, we expect exponential convergence with increasing number of modes. 

\subsection{Singular solutions}
\label{app:numerics_singular}
We next assess the suitability of the above ``straightforward'' scheme in the case where the solution exhibits the equatorial blowup singularity of the form identified by local analysis in Sec.~\ref{sec:local}. Using the Taylor symmetry, the Fourier inversion of \eqref{q fourier} can be written
\begin{equation}\label{alpha inversion}
\alpha_n = \frac{4}{\pi}\int_0^{\pi/2}q\cos((2n-1)\theta)\,\mathrm{d}\theta, \quad n=1,2,\ldots
\end{equation}
Noting the local behaviour (\ref{blow up}a), it is clear that for large $n$ the integrals on the right-hand side of \eqref{alpha inversion} are dominated by a local contribution near $\theta=\pi/2$, 
\begin{multline}
\alpha_n\sim -\frac{4A}{\pi}\int_{-\infty}^{\pi/2}(\pi/2-\theta)^{-1/3}\cos((2n-1)\theta)\,\mathrm{d}\theta\\=(-1)^n\frac{4A}{\pi}\int_0^{\infty}s^{-1/3}\sin((2n-1)s)\,\mathrm{d}s \quad \text{as} \quad n\to\infty,
\end{multline}
or
\begin{equation}\label{alpha hat}
\alpha_n\sim A \hat{\alpha}_n \quad \text{as} \quad n\to\infty, \quad \text{where} \quad \hat{\alpha}_n=\frac{4}{2^{2/3}\Gamma(1/3)}(-1)^{n}n^{-2/3},
\end{equation}
in which $A$ is the singularity prefactor appearing in the local expansions \eqref{blow up} and $\Gamma(z)$ denotes the Gamma function of argument $z$.
Similarly, the large-$n$ behaviour of $\chi_n$ can be obtained by inverting (\ref{v and stress fourier}a) using the local behaviour (\ref{blow up}b). In this case, since $\tilde{v}$ is weakly singular, the Fourier integrals must be integrated by parts once before the leading-order behaviour can be extracted as a local contribution. We thereby find
\begin{equation}\label{chi hat}
\chi_n\sim A^2 \hat{\chi}_n \quad \text{as} \quad n\to\infty, \quad \text{where} \quad \hat{\chi}_n = \frac{4}{2^{1/3}\Gamma(2/3)}(-1)^{n+1}n^{-1/3}.
\end{equation}

With the asymptotic behaviours \eqref{alpha hat} and \eqref{chi hat}, we see that the series \eqref{q fourier} for $q$ converges conditionally, except at the equator, $\theta=\pm\pi/2$, where it vanishes identically. As the equator is approached --- in which limit $q$ itself diverges --- the convergence deteriorates, with the $n^{-2/3}$ attenuation of the terms in the series predicted by \eqref{alpha hat} only apparent for $n$ of order $1/|\theta\pm\pi/2|$ or larger. Accordingly, the Fourier series for $q$ is useless for numerical evaluation owing to the Gibbs phenomenon. In contrast, the series (\ref{v and stress fourier}a) for $\tilde{v}$ and \eqref{qv fourier} for $q\tilde{v}$, which are non-smooth but whose two-sided limits exist at the equator, converge absolutely for all $\theta$. 
{In principle, $q$ can be evaluated by dividing  $q\tilde{v}$ and $\tilde{v}$, however with the present scheme this procedure is impractical owing to the slow convergence of the corresponding Fourier series --- the quotient is particularly sensitive to errors near the poles $\theta=0,\pi$ where both $q\tilde{v}$ and $\tilde{v}$ are small.

Let us briefly comment also on the applicability of the straightforward scheme in the case where the solution exhibits a weaker equatorial singularity, of the form observed in the transition region described in Sec.~\ref{ssec:transition}. In that scenario, $q$ is continuous, exhibiting near the equator the sublinear behavior \eqref{sublinear}, with exponent $0<\alpha<1$, whereas $\tilde{v}$ scales linearly with $\pi/2-\theta$ as $\theta\to\pi/2$, with an order $|\pi/2-\theta|^{1+\alpha}$ correction. Following the same approach as above, we find that the Fourier coefficients $\alpha_n$ and $\chi_n$ now decay like $1/n^{1+\alpha}$ and $1/n^{2+\alpha}$, respectively, hence the Fourier series for $q$ and $\tilde{v}$ converge absolutely. Clearly, computing $q$ in this way becomes impractical for sufficiently small $\alpha$. The Fourier series for $d\tilde{v}/d\theta$ also exhibits a slow decay rate, as $1/n^{1+\alpha}$, which for sufficiently small $\alpha$ prohibits the calculation of the exponent $\alpha$ via \eqref{exponent}. 

\subsection{Singularity-capturing scheme}
\label{app:numerics_improved}
We next develop a modified singularity-capturing scheme tailored to calculating steady solutions exhibiting equatorial blowup. A conventional approach would be to subtract the known local behaviour of the solution in physical space \cite{Peyret:02}, namely \eqref{blow up} in the present problem. This has the disadvantage of adding inhomogeneous terms to the governing equations, including the Laplace and biharmonic equations governing the bulk potential and stream function, meaning that we would have to abandon the simple general solutions \eqref{pot fourier} and \eqref{psi fourier}. Instead, our approach will be to account for the local behaviours \eqref{blow up} directly in Fourier space. Specifically, we choose a large integer $N$ and set $\alpha_n=A\hat{\alpha}_n$ and $\chi_n=A^2\hat{\chi}_n$ for $n>N$ [cf.~\eqref{alpha hat} and \eqref{chi hat}]. Then, instead of truncating the infinite sums in \eqref{gamma coefficients}, we find, for $n=1,2,\ldots,N$, 
\begin{equation}\label{gamma coefficients A}
\gamma_n=G_n+AH_n+A^2I_n+A^3J_n, \quad 
\end{equation}
wherein
\begin{equation}
G_n=\sum_{k=1}^{n-1}\frac{\chi_k\alpha_{n-k}}{8k}-\sum_{k=1}^{N-n}\frac{\chi_{k}\alpha_{n+k}}{8k}+\sum_{k=1}^{N-n+1}\frac{\chi_{n+k-1}\alpha_k}{8(n+k-1)},
\end{equation}
\refstepcounter{equation}
$$
H_n=\sum_{k=N-n+1}^N\frac{3\hat{\alpha}_{n+k}}{8k}\chi_k, \quad I_n=\sum_{k=N-n+2}^N\frac{9\hat{\chi}_{n+k-1}}{8(n+k-1)}\alpha_k
\eqno{(\theequation \mathrm{a},\mathrm{b})}
$$
and
\begin{equation}
J_n=\frac{27}{8}\sum_{k=N+1}^{\infty}\left\{\frac{\hat{\chi}_k\hat{\alpha}_{n+k}}{k}-\frac{\hat{\chi}_{n+k-1}\hat{\alpha}_k}{n+k-1}\right\}.
\end{equation}
We note that the sums $J_n$ are absolutely convergent and can be calculated \textit{a priori} in a pre-processing stage. The unknowns are the coefficients $\alpha_n$ and $\chi_n$, for $n=1,2,\ldots,N$, as well as the singularity prefactor $A$. The governing equations are \eqref{chi coefficients} and \eqref{qv condition fourier}, with \eqref{gamma coefficients A}, for $n=1,2,\ldots,N$, as well as the global relation \eqref{A in Q}, which connects $A$ to $\mathcal{Q}$ and so to the $\alpha_n$ coefficients via \eqref{quadrant charge fourier}. We employ the Matlab solver \texttt{fsolve} \cite{MATLAB:R2022b} with the Jacobian provided analytically. 

There are two ways to evaluate the solutions. The straightforward approach is to evaluate the Fourier-series representations for $n=1,2,\ldots,N'$, for some $N'\gg N$ (with $q$ and $\partial\tilde{v}^-/\partial{r}$ evaluated indirectly as explained in subsection  \ref{app:numerics_singular}). Alternatively, the large-$n$ tail contributions can be summed analytically. 
For example, we can re-write the Fourier series (\ref{v and stress fourier}a) for the tangential component of the surface velocity as 
\begin{equation}\label{v decomposition}
\tilde{v}=A^2\mathcal{V}(\theta)+\sum_{n=1}^{N}\left(\frac{\chi_n}{4n}-A^2\frac{\hat{\chi}_n}{4n}\right)\sin 2n\theta, 
\end{equation}
where $A$ and the finite sum are evaluated numerically (the terms in the sum oscillate and decay faster than $1/n^{4/3}$), while 
\begin{equation}\label{V analytical}
\mathcal{V}(\theta)= \sum_{n=1}^{\infty}\frac{\hat{\chi}_n}{4n}\sin 2n\theta=-\frac{i}{2^{4/3}\Gamma(2/3)}\left\{\mathrm{Li}_{4/3}\left(-e^{-2i\theta}\right)-\mathrm{Li}_{4/3}\left(-e^{2i\theta}\right)\right\},
\end{equation}
in which $\mathrm{Li}_{4/3}(z)$ denotes the polylogarithm function of order $4/3$ and argument $z$, with $i=\sqrt{-1}$. Using an integral representation of the latter function \cite{Abramowitz:book}, we can express $\mathcal{V}$ as a real integral,
\begin{equation}
\mathcal{V}(\theta)
=\frac{3^{3/2}}{2^{1/3}4\pi}\int_0^{\infty}\frac{s^{1/3}\sin 2\theta}{\cosh s + \cos 2\theta}\,\mathrm{d}s.
\end{equation}
In \eqref{v decomposition}, the first term is an analytical function that accounts for the leading local behaviours of $\tilde{v}$ as $\theta\to\pm\pi/2$, whereas the second term converges as $N\to\infty$ to a profile that is smoother than $\tilde{v}$. We can similarly regularise other non-smooth surface profiles. Furthermore, the tangential velocity profile $\mathcal{V}(\theta)$ can be extended to a homogeneous solution of the flow problem that could in principle be used to subtract the leading local behaviour of the velocity field in physical space without introducing forcing terms into the governing equations. We could similarly produce suitable non-smooth homogeneous solutions of the potential problem. 

\section{Condition on local behaviour near cap edge} 
\label{app:regularity}
In this appendix, we derive the regularity condition \eqref{cap regularity} included in the electric cap problem of Sec.~\ref{ssec:cap_electric} by considering the local behavior of the  interior potential $\tilde{\varphi}^-_0$ near the cap edge $(r,\theta)=(1,\theta^*)$. Let $(\varrho,\vartheta)$ be polar coordinates from that point such that $\varrho$ is the distance from the cap edge and $\vartheta$ the angle measured counter-clockwise from the $\be_{\theta}$ direction at $\theta^*$, see Fig.~\ref{fig:cap_edge_local}. To leading order in the local limit $\varrho\searrow0$, 
\refstepcounter{equation}
$$
\label{cap local coordinates}
\theta-\theta^*=\varrho\cos\vartheta, \quad 1-r=\varrho\sin\vartheta,
\eqno{(\theequation \mathrm{a},\mathrm{b})}
$$
and the interface is locally flat, such that $\mathcal{C}$ coincides with $\vartheta=\pi$ and $\overline{\mathcal{C}}$ with $\vartheta=0$. 

To leading order as $\varrho\searrow0$, the potential perturbation $\varphi'=\tilde{\varphi}_0^-+\tilde{\varphi}^*$ satisfies Laplace's equation for $\vartheta \in (0,\pi)$, together with the inhomogeneous Neumann boundary condition [cf.~\eqref{cap q0 bc explicit}]
\begin{equation}\label{cap edge inhomogeneous}
\pd{\varphi'}{\vartheta}=2\varrho \cos\theta^* \quad \text{at} \quad \vartheta=0
\end{equation}
and the Dirichlet condition 
\begin{equation}\label{cap edge Dirichlet}
\varphi'=0 \quad \text{at} \quad \vartheta=\pi.
\end{equation}
Also to leading order as $\varrho\searrow0$, the perturbation of the charge density $q_0$ from $2\cos\theta$ is [cf.~\eqref{q0 and phi0}]
\begin{equation}
q_0 -2\cos\theta= \mp\frac{1}{\varrho}\pd{\varphi'}{\vartheta}  \quad \text{for} \quad \vartheta=0,\pi,
\label{edge q perturbation}
\end{equation}
respectively, 
where note that $\cos\theta\sim \cos\theta^*$ with an $\mathcal{O}(\varrho)$ error, which we shall see is negligible relative to the right-hand side of \eqref{edge q perturbation}. 
\begin{figure}[t!]
\begin{center}
\includegraphics[scale=0.7,trim={0 1cm 0 1cm}]{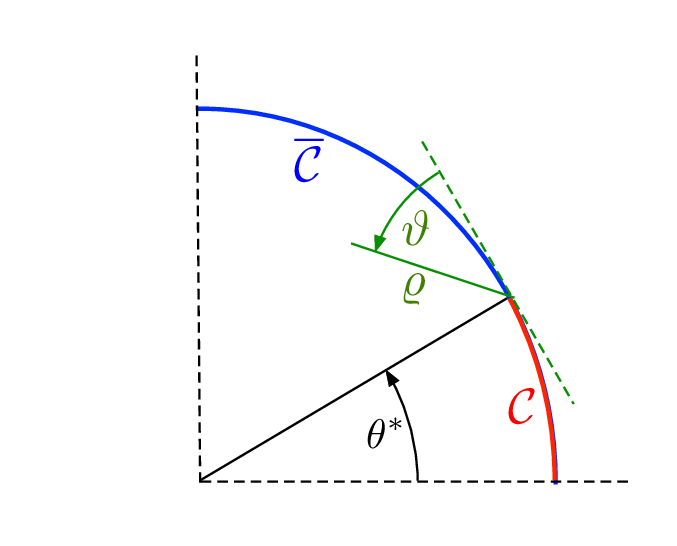}
\caption{Local cap-edge polar coordinates $(\varrho,\vartheta)$ used in Appendix \ref{app:regularity}.}
\label{fig:cap_edge_local}
\end{center}
\end{figure}

The inhomogeneous boundary condition \eqref{cap edge inhomogeneous} suggests an order-$\varrho$ potential perturbation, 
\begin{equation}\label{cap edge linear}
\varphi' \sim  2\cos\theta^*\varrho\sin\vartheta \quad \text{as} \quad \varrho\searrow0,
\end{equation}
such that \eqref{edge q perturbation} gives $q_0=o(1)$ at both $\vartheta=0$ and $\pi$. However, we must also consider homogeneous local solutions of Laplace's equation and the boundary conditions \eqref{cap edge inhomogeneous} and \eqref{cap edge Dirichlet}, which may, in fact, dominate the perturbation \eqref{cap edge linear}. Indeed, such homogeneous solutions are
\begin{equation}
\label{homo solutions} \varrho^{1/2+n}\cos[(1/2+n)\vartheta],
\end{equation} 
where $n$ is an integer that must be non-negative so that $q_0$ remains integrable. Thus, rather than \eqref{cap edge linear}, we may generally expect the local behavior 
\begin{equation}
\varphi'\sim \text{const.}\times \varrho^{1/2}\cos\frac{\vartheta}{2} \quad \text{as} \quad \varrho\searrow0,
\label{edge wrong local}
\end{equation} 
implying that $q_0$ diverges like $1/\varrho^{1/2}$ as $\varrho\searrow0$ on $\mathcal{C}$, i.e., as $\theta\nearrow\theta^*$. 
Recalling that $q_0$ represents an outer approximation, such divergence implies that $q$ is asymptotically large in an inner region near $\theta=\theta^*$, which seems implausible; we expect such an inner region to enable a smooth transition of $q$ from order unity values on $\mathcal{C}$ to order $1/\widetilde{\mathrm{Re}}^{1/2}$ values on $\overline{\mathcal{C}}$. While an analysis of the inner region is beyond the scope of this paper, this variation between order unity and $o(1)$ values suggests that the homogeneous solution \eqref{edge wrong local}, corresponding to $n=0$ in \eqref{homo solutions}, must be eliminated, leaving the smoother leading-order behaviour \eqref{cap edge linear},  with an $\mathcal{O}(\varrho^{3/2})$ error associated with the $n=1$ homogeneous solution \eqref{homo solutions}. The associated charge density is therefore $\mathcal{O}(\varrho^{1/2})$. It follows that the outer approximations $q_0$ and $\partial\tilde{\varphi}^-_0/\partial\theta$ are continuous functions at $r=1$, giving the regularity condition \eqref{cap regularity} assumed in the main text.  

\bibliography{refs}
\end{document}